\documentstyle[11pt,draft,rotate,epsf,lprocl]{article}

\newcommand{\be}{\begin{equation}}
\newcommand{\ee}{\end{equation}}

\newcommand{\bea}{\begin{eqnarray}}
\newcommand{\eea}{\end{eqnarray}}

\newcommand{\bfi}{\begin{figure}}
\newcommand{\efi}{\end{figure}}

\newcommand{\cH}{{\cal H}}
\newcommand{\cZ}{{\cal Z}}

\newcommand{\overl}[1]{\overline{#1}}
\newcommand{\lan}{\langle}
\newcommand{\ran}{\rangle}
\newcommand{\sign}{\mbox{ sign}}
\newcommand{\qea}{q_{\rm EA}}

\newcommand{\al}{\alpha}
\newcommand{\si}{\sigma}
\newcommand{\eps}{\epsilon}
\newcommand{\ga}{\gamma}

\begin{document}

\title{NUMERICAL SIMULATIONS OF SPIN GLASS SYSTEMS}
\author{E. MARINARI}
\address{Dipartimento di Fisica and INFN, Universit\`a di Cagliari,
Via Ospedale 72,\\
09100 Cagliari, Italy}

\author{G. PARISI and  J.J. RUIZ-LORENZO}
\address{Dipartimento di Fisica and INFN, Universit\`a di Roma ``La 
Sapienza'',
P. A. Moro 2\\
00185 Roma, Italy}

\maketitle

\abstracts{
We discuss the status of Monte Carlo simulations of (mainly finite 
dimensional) spin glass systems. After a short historical note and a 
brief theoretical introduction we start by discussing the (crucial) 
$3D$ case: the warm phase, the critical point and the cold phase, the 
ultrametric structure and the out of equilibrium dynamics. With the 
same style we discuss the cases of $4D$ and $2D$. In a few 
appendices we give some details about the definition of states and 
about the tempering Monte Carlo approach.
}

\section{Introduction\protect\label{S-INTROD}}

Spin glasses are a fascinating subject, both from the experimental and 
from the theoretical point of view \cite{BINYOU,MEPAVI,FISHER,PARISI}.  
In the framework of the mean field approximation a deep and complex 
theoretical analysis is needed to study the infinite range version of 
the model (the Sherrington-Kirkpatrick model, SK model in the 
following).  Using the formalism of replica symmetry breaking 
\cite{BREAKS} (RSB) one finds an infinite number of pure equilibrium 
states, which are organized in an ultrametric tree.  It is fair to say 
that while most of the equilibrium properties of the SK model are well 
understood, much less is known about the detailed features of the 
dynamics, although recent progresses have been done in this direction.

A crucial question is how much of this very interesting structure 
survives in short range models, defined in finite dimensional space.  
Numerical simulations are very useful for trying to answer this 
question, since most of the more peculiar predictions are for 
quantities that is difficult to relate to measurements that can be 
performed in real experiments. 

Our goal will eventually be to draw a meaningful comparison of the 
theoretical findings and the experimental data.  In order to do that 
we will discuss the mean field picture that we have introduced before 
and a different point of view, the droplet model 
\cite{MCMILL,BRAMOO,FISHUS}.  We will see that a comparison of the 
predictions of the mean field theory with those arising from the 
droplet model systematically shows the appropriateness of the mean 
field picture.

In the most part of cases an interacting theory is formulated by 
starting from a limiting case which is well under control.  Then one 
constructs some kind of perturbation expansion, but the features one 
finds in this way typically share many features with the starting 
point one used: one better starts from a good guess.  In spin glasses 
there are two different starting point that have been considered in 
the literature:

\begin{itemize}

\item
The mean field approximation, which is correct in the infinite
dimensional limit.

\item
The Migdal-Kadanoff (MK) approximation \cite{MIGKAD}, which is 
(trivially) correct in one dimension and for some fractal lattices 
(e.g.  carpet lattices).  This approximation is the basis of the so 
called droplet model (hereafter DM).

\end{itemize}

It is known that the MK approximation gives results that are violently 
wrong from a quantitative point of view when we go to a large 
dimensionality space (in most models the results are acceptable only 
in dimension $2$ or less).  For example in a ferromagnet the MK 
approach does not detect the triviality of the critical exponents in 
dimensions greater than $4$.  Usually the MK approximation grasps 
correctly the qualitative behavior (e.g.  the existence of Goldstone 
modes in models with spontaneously broken $O(N)$ symmetry) in the low 
temperature phase and from this point of view it agrees with the mean 
field predictions.

There 
is no controversy on the behavior at the transition point in spin glasses
is concerned in zero external magnetic field.  Critical exponents are given 
by the mean field in more than six dimension and a (poorly convergent) 
$\epsilon$-expansion predicts the exponents in $6-\epsilon$ dimensions 
\cite{GREEN85,HALUCHE76}.

On the contrary in the low temperature phase the two approaches imply 
a very different behavior.  Mean field theory predicts that for a 
large, finite system \footnote{We discuss the problem of defining a 
state in the finite volume spin glass system in Appendix 
(\ref{S-PURE}).}, there are many different equilibrium states.  The 
droplet approach predicts that the equilibrium state is unique, apart 
from reflections.  The two points of view drastically differs in the 
properties of overlap: in the droplet model the value of the overlap 
$q$ among two different real replicas of the systems is expected to be 
a given number, while in the mean field approach it has a non trivial 
probability distribution $P(q)$, which in the infinite volume limit 
has support in the interval $(q_m,q_M)$ ($q_m$ stands for the minimum 
$q$ value, $q_M$ stand for the maximum $q$ value).  The value $q_M$ 
coincides with the overlap among two generic configurations in the 
same state, which is denoted $\qea$ ($\rm EA$ stands here for 
Edwards-Anderson). The probability distribution for a given sample 
$P_J(q)$ is a quantity that depends on the sample: it is a non 
self-averaging quantity.

This difference in the expectations for $q$ has strong implications for the 
magnetic susceptibility: in the droplet model in the limit of zero magnetic 
field there is no ambiguity in the definition of the susceptibility and it  is 
given by the relation

\be \chi =  \beta (1-\qea)\ . \ee 

In the mean field approach there are two different susceptibilities:
\begin{itemize}
\item The linear response susceptibility ($\chi_{LR}$) which is given by 
the zero frequency limit of the time dependent susceptibility (equivalently it 
is given by variation of the magnetization when an infinitesimal magnetic field 
is applied to a system in a pure state).  It is given by:
\be \chi_{LR} = \beta (1-\qea)\ .  \ee

\item The equilibrium susceptibility, i.e.  the derivative of the equilibrium 
magnetization with respect to the magnetic field.  It is given by the relation:

\begin{equation} 
\chi_{eq} = \beta\int {\rm d}q~(1-q) P(q) \ge \chi_{LR} \ .
  \protect\label{E-CHIROTTA}
\ee 

With very good approximation $\chi_{eq}$ is given experimemtally by the 
derivative of the thermoremanent magnetization with respect to the magnetic 
field.
\end{itemize}

In the droplet model the two suscetibilities are equal.  In the mean field 
approach we have $\chi_{LR}<\chi_{eq}$ in the broken phase, while in the warm 
phase $\qea=0$ and we get that both susceptibilities are given $\beta$.  In the 
broken phase region we have $\chi_{eq}>\beta (1-\qea)$.

 The difference of the two susceptibilities is a typical prediction of the 
mean field theory; indeed in the first case ($\chi_{LR}$) the system in presence 
of an infinitesimal magnetic field must be very similar from that in zero 
magnetic field.  In the second case systems at equilibrium in different magnetic 
fields may correspond do very different microscopical configurations.

 In many cases numerical simulations 
have been extremely useful to discriminate among different theoretical 
scenarios and to discover the existence of possible non perturbative 
effects.  Spin glasses are not an exception to this rule, although 
numerical simulations are much more difficult here than in the usual 
ferromagnetic case \cite{RIEGER95}.  The main difficulty is related to 
the high value of the dynamic exponent $z$.  Already in mean field 
$z$ is quite large ($4$) and it becomes still larger in three 
dimensions (around $6$).  This is very different from usual 
ferromagnets, where $z$ has a small value (close to $2$), largely 
independent from the system dimensionality.

Most of the numerical simulations have been ran in three dimensions, 
where it is more difficult to get satisfactory results (we will 
discuss this issue in much detail in the following).  The situation in 
two and four dimension has been clarified by numerical simulations 
(for opposite reasons: see later) in a far more complete and 
satisfactory way.  Although the behavior of finite dimensional spin 
glass systems in presence of a magnetic field is very interesting
unfortunately only few data are available.

In section (\ref{S-HYSTOR}) we use a few phrases to describe the 
earlier generation series of Monte Carlo simulation: we will not have 
space to describe them in detail, and we will just draw the main 
findings.  In section (\ref{S-DEFINI}) we define the models, and give 
the definitions we will use in the text.  In section (\ref{S-MINI}) we 
give a mini-theoretical review.  We start the bulk of our discussion 
by the crucial case of $3$ dimensions (\ref{S-D3}): we discuss 
simulations in the high $T$ phase (\ref{SS-D3WARM}), in the broken 
phase (\ref{SS-D3COLD}), simulations using three replicas of the 
system (\ref{SS-D3THRE}) and off-equilibrium dynamic simulations 
(\ref{SS-D3DYNA}).  We discuss how the existence of a phase transition 
has been made clear, and how one qualifies the broken phase, showing 
it is broken according to the mean field RSB pattern.  After that we 
discuss the case of $4D$ (\ref{S-D4}), where the existence of a mean 
field like broken phase it is absolutely clear from the numerical 
point of view.  The case of $2D$ , where one does not have a finite 
$T$ phase transition, is discussed in (\ref{S-D2}) to stress peculiar 
effects and behaviors of interest.  In a series of appendices we 
discuss about pure state (\ref{S-PURE}), and about improved Monte 
Carlo Methods (tempering (\ref{S-TEMPER}) and parallel tempering 
(\ref{S-PARTEM})).

We realize that there are many very interesting subject that we have 
not considered for lack of space: we only quote the numerical 
simulations in Hamiltonian infinite range models with Ising, 
Heisenberg or spherical spins with interactions connecting two or more 
spins\cite{BJMN,BA,STA}; the whole series of questions connected to 
non-Hamiltonian system \cite{IORMAR}; non-Ising spins in finite dimensions 
\cite{COLUZZI95}; Ising spin glass at the upper critical dimension 
\cite{WAYO93}; chaos in spin glasses \cite{RITORT} and quantum spin 
glasses \cite{QSG}.  There is surely much more that we are omitting, and we 
apologize.

\section{History\protect\label{S-HYSTOR}}

The papers by Ogielski and Morgenstern \cite{OGMO85} and by Bhatt and 
Young \cite{BHAYO85} start somehow the history of modern, large scale 
simulations of finite dimensional spin glass models.  They both deal 
with $3D$ systems, with quenched random couplings $J=\pm 1$ with 
probability $\frac12$.  A special purpose computer has been built for 
running the simulations of \cite{OGMO85}: this has been one of the 
milestones of the history of computers dedicated or optimized (as far 
as the hardware is concerned) for the study of problems in theoretical 
physics.

Ref.  \cite{OGMO85} deals with both equilibrium and dynamics.  The
best output of the simulations is that there is a phase transition at
$T_c=1.20\pm 0.05$, with $\nu=1.2\pm 0.1$, but if a $T=0$ power law
divergence can be excluded an exponential divergence of the kind
$\xi\simeq\exp(b/T^c)$ (that is what we expect at the
lower critical dimension, LCD, see later) fits very well the
data. The dynamic simulations allow to estimate a correlation time
that assuming a phase transition scales like $\tau(T)\simeq\xi^z$,
with $z\simeq 5$. An exponential fit to a LCD form works fine. Also a
Vogel-Fulcher behavior $\tau\simeq\tau_0\exp( \frac{\Delta F}{T-T_0})$
with $T_0\simeq 0.9$ fits well the data.

If one assumes the existence of a phase transition the work of 
\cite{BHAYO85} gives compatible results, with $T_c\simeq 1.2$, 
$\nu=1.3\pm 0.3$ and $\eta = -0.3 \pm 0.2$, but the simulations at 
not so high $T$ values make the possible LCD behavior very clear. 
The spin glass susceptibility $\chi_q$ is estimated here with two 
different approaches (two copies of the system or dynamic correlation 
functions). The two possibilities of $3$ being the LCD and of a 
Kosterlitz-Thouless like transition are compatible with the data.

In a longer paper Ogielski \cite{OGIELSKI85} mainly discusses the 
dynamic behavior of the $3D$ system. He finds that for $T>T_c$ the 
dynamic correlation functions can be described by a stretched 
exponential decay, while in the cold phase one always detects power 
law (a typical signature of the slow dynamics of a complex system). 
The dynamic exponent $z$ turns out to be close to $6$. Again, one 
gets hints for dimension $3$ being marginal or close to it. The 
dynamic behavior of the $3D$ model has also been studied by Sourlas 
\cite{SOURLA}, while looking at domain walls gives compatible 
results \cite{OGIELW}.

Bhatt and Young \cite{BHAYO88} study the cases $2D$, $3D$ and $4D$, 
with a systematic analysis of the Binder parameter $g$ (and an 
accurate study of thermalization).  In $2D$, with $J=\pm 1$ (here 
there can be a difference from the case of continuous couplings, since 
the ground state has an accidental degeneracy: $\eta$ for example is 
not expected to be universal) $T_c=0$.  Assuming a power divergence 
gives $\nu=2.6\pm 0.4$, $\eta=.20\pm .05$ and $\gamma = 4.6 \pm 0.5$.  
In $3D$ they study the case of Gaussian couplings, to investigate 
universality.  Again one finds that the existence of a phase 
transition is favored, but the LCD is very close.  $4D$ appears as an 
easy case.  The critical region is clear, and one can easily get a 
rough but reliable estimate $\nu=0.80\pm 0.15$, $\eta=-.30\pm .15$ and 
$\gamma = 1.8 \pm 0.4$.

Ergodicity breaking in $3D$ has been discussed by Sourlas in
\cite{SOURLB}. 

The work by Reger, Bhatt and Young \cite{REBHYO} uses the observation 
that $4D$ is a simple case to make it a test case.  Ref.  
\cite{REBHYO} clearly shows that the broken phase of the $4D$ system 
has a non-trivial overlap probability distribution: things go exactly 
as they do in the mean field model.  After ref.  \cite{REBHYO} one has 
to turn a triple somersault in order to claim that the mean field 
limit is not a good starting point to study the realistic case of 
finite $D$ dimensional models, with $D$ lower than the upper critical 
dimension and higher than the lower one.

\section{Definitions\protect\label{S-DEFINI}} 

We give here some definitions that will be needed in the following.  We work in 
$D$ spatial dimensions.  The linear extension of our lattice is $L$, and the 
volume is $V=L^D$ (sometimes we will denote it with $N$).  In the mean field 
model $N$ or $V$ denote the total number of lattice sites.  Typically we work 
with Ising spins $\sigma_i=\pm 1$.  The Hamiltonian is
\be H \equiv \sum_{<i,j>}\sigma_i J_{i,j} \sigma_j\ , \ee 
where 
the sum runs over first neighbor on the $D$ dimension (simple cubic, where we do 
not specify something different) lattice, and the $J$ are quenched random 
variables.  The couplings $J$ will be sometimes Gaussian, and sometimes they 
will take the value $\pm 1$ with probability $\frac12$ (see text).  The 
magnetization is

\be
  m \equiv \frac{1}{V}\sum_i \sigma_i \ .
\ee
In spin glasses it is not a very interesting quantity, since
by using the gauge invariance of the theory one can show that  
$\overl{\langle m\rangle}=0$. The magnetic susceptibility is

\be
  \chi \equiv \frac{1}{V} 
  \overl{ \lan m^2\ran } \ . 
\ee
The overlap among two configurations $\alpha$ and $\beta$ at site $i$is

\be
  q_i^{\alpha,\beta}
  \equiv \sigma_i^{\alpha}   \sigma_i^{\beta}\ ,
  \protect\label{E-QI}
\ee
and the total overlap

\be
  q^{\alpha,\beta}\equiv\frac{1}{V}\sum_i q_i^{\alpha,\beta}\ ,
  \protect\label{E-QSUM}
\ee
where we will frequently ignore the superscripts by denoting it with 
$q$. The overlap is the essential ingredient for the study of a spin 
glass. Its probability distribution for a given sample is

\be 
  P_J(q^\prime)=\lan \delta(q^\prime-q)\ran\ ,
\ee 
and averaging over samples one has

\be 
  P(q) \equiv \overl{P_J(q)}\ .  
  \protect\label{E-PQ}
\ee 
The Binder parameter has a crucial role in locating phase transitions:

\be
  g \equiv \frac12 
  \left[
  3-\frac{\overline{\langle q^4\rangle}}{\overline{\langle 
  q^2\rangle}^2} \right] \ .
  \label{E-BINDER}
\ee
It scales as

\be
  g = \tilde{g}
  \left(
  L^{\frac{1}{\nu}}
  \left(T-T_c\right)
  \right)\ ,
  \label{E-BINSCA}
\ee
i.e. at $T_c$ the Binder parameter does not depend on $L$ 
(asymptotically for large $L$ values). In some parts of the text we will also 
denote it by $B$. The overlap susceptibility is defined as

\be
  \chi_q \equiv \lim_{V\to\infty}V  \overline{\langle q^2\rangle} \ .
  \protect\label{E-CHIQ}
\ee
The spatial overlap-overlap correlation function is

\be
  G_{i,j} \equiv \overline{
  \langle q_i q_{i+j} \rangle}
  =  \overline{
  \langle \sigma_i \tau_i \sigma_{i+j} \tau_{i+j}\rangle}
  =  \overline{
  \langle \sigma_i \sigma_{i+j}\rangle^2}\ ,
  \protect\label{E-GIJ}
\ee
and

\be
  G_j \equiv \frac{1}{V}\sum_i G_{i,j}\ .
  \protect\label{E-GJ}
\ee
Sometimes we will indicate $G_j$ with $G(j)$ or $G(x)$.

In our numerical simulations we measure sometimes the non connected 
overlap-overlap correlation function
\be
  G^{(\cdot)}(d)
  \equiv \sum_{i,j=(i+d,0,0)}G_{i,j}\ ,
  \protect\label{E-GSISI}
\ee
where the sum runs in a single, given direction of the lattice.  From 
here one can for example define an effective distance dependent 
correlation length

\be
  \tilde{\xi}^{(\cdot)}(d) \equiv
  \log\left( \frac{G^{(\cdot)}(d+1)}{G^{(\cdot)}(d)}
  \right)\ ,
  \protect\label{E-CSISI}
\ee
that for $d\to\infty$ tends to the asymptotic correlation length.
The connected overlap-overlap correlation function is defined as

\be
  \widehat{G}_j^{(q)} \equiv G_j - q^2\ .
  \protect\label{E-GCON}
\ee
We have made explicit the dependence of $\widehat{G}_j^{(q)}$ over 
$q$: one can select states with a given overlap $q$ and compute the 
correlation among them.

At last an important tool to study the dynamic of a system is the spin-spin
autocorrelation function, i.e.

\be
  C(t,t_w) \equiv
  \frac{1}{V} \sum_{i=1}^V
  \overl{\langle \sigma_i(t_w) \sigma_i(t_w+t) \rangle}\ .
  \protect\label{E-CORDYN}
\ee

\section{A Mini-Theoretical Review\protect\label{S-MINI}}

The aim of this section is to recall the predictions of the mean field
approximation and to clarify the language we are using in the rest of
the paper.

\subsection{Some Mean Field Useful Results\protect\label{SS-USEFUL}}

In the mean field theory the probability distribution of the overlaps 
averaged over the disorder, (\ref{E-PQ}), has a smooth part plus a 
delta function at $\qea$.  We have already said that the function 
$P_J(q)$ fluctuates with the coupling realization $J$.  In the replica 
formalism \cite{MEPAVI} one finds that

\be 
  \overl{P_J(q_1)P_J(q_2)} =
  \frac13 P(q_1)\delta(q_1-q_2)+\frac23 P(q_1) 
  P(q_2)\ .
  \label{E-GUERRA}
\ee
This relation tells us something about the fluctuations of the 
function $P_J(q)$.  It has been recently proven rigorously by Guerra 
under very general assumptions \cite{GUERRA95}.  {\em Ultrametricity} 
\cite{RATOVI} is another very interesting property, which we will 
discuss in detail in sections (\ref{SS-D3THRE}) and (\ref{SS-D4ULTR}).

A crucial property of the pure states is the vanishing at large 
distance of the connected correlation function (\ref{E-GCON}) among 
two states $\al$ and $\ga$.  It is also evident that the correlation 
$\widehat{G}_x^{(q)}$ (\ref{E-GJ}) depends on $q$. Its value at $j=1$ 
is particularly interesting, and in the case of the models with $J=\pm 
1$ it is equal to the average of the so-called energy overlap:

\be
  q_e \equiv \frac{1}{V}
  \sum_y \overl{\lan\si_{y+1}J_{y,y+1}\si_y\ran_\al
  \lan\si_{y+1}J_{y,y+1}\si_y\ran_\ga} \ .
\ee
Also the asymptotic behavior of the function $\widehat{G}_x^{(q)}$ for 
large $x$ is interesting.  By a tree level computation the authors of 
\cite{DOKO86} find that

\bea
\widehat{G}_x^{(q)} \propto \left\{ \begin{array}{ll}
                      x^{-D+2}   & \mbox{ if } q=q_{\rm EA}\ ,\\
                      x^{-D+3}   & \mbox{ if } 0<q<q_{\rm EA}\ ,\\
                      x^{-D+4}   & \mbox{ if } q=0\ .
                      \end{array}\right.
\eea
These predictions are valid close to the upper critical dimension, 
$6$, and they will surely be modified in a number of dimensions small 
enough.
In particular a systematic perturbation theory \cite{DOKO93} gives indications 
that in less than $6$ dimensions

\be
  \widehat{G}_x^{(q=0)}\simeq x^{{-D+2-\eta}\over 2}\ ,
\ee
where $\eta$ is the usual critical exponent computed at the phase 
transition point.  The function $\widehat{G}_x^{(q=0)}$ is interesting also 
because it the most accessible by numerical simulations: one does not 
need to fix the constraint, but it can be automatically implementing by starting 
with two non thermalized configurations on a large lattice.  In such a situation 
the system will stay in the $q=0$ sector for a very large time (more precisely 
for a time which diverges when the volume goes to infinity), since the two copies 
will typically approach thermal equilibrium by relaxing in two orthogonal 
valleys.

The existence of a whole set of $q$-dependent correlation functions 
with different critical behaviors is a crucial prediction of the mean 
field theory. The usual overlap correlation functions which are 
obtained by integrating over the whole phase space are given by 

\be
  \widehat{G}_x\equiv \int {\rm d}q\ P(q)\ \widehat{G}_x^{(q)}\ .
  \protect\label{E-GALL}
\ee  
These features are not shared by the droplet model.  In the DM the 
function $P_J(q)$ always contains a single delta function (two at zero 
magnetic field, $h=0$, because of the spin reversal symmetry): if at 
$h=0$ we consider only one of each couple of states obtained by 
changing the sign of all the spins of the lattice DM tells us that the 
system has only one state.

\subsection{Coupled Replicas\protect\label{SS-COUPLE}}
 
The introduction of an interaction among replicas 
\cite{FRPAVA,CAPAPASO90} (i.e.  different spin configurations which 
are defined in the same quenched couplings) generates a very interesting 
phenomenology.  Let us consider a system of two 
replicas $\si$ and $\tau$, described by the Hamiltonian

\be 
  H_{J}(\si,\tau) \equiv 
  H_{J}(\si)+H_{J}(\tau)- \eps\sum_{i}\si_{i}\tau_{i} \  .
\protect\label{EPSQ}
\ee
In the mean field theory one finds that the expectation value of the 
overlap $q$ among the two replicas $\sigma$ and $\tau$ for small 
$\eps$ behaves as

\be
  q(\epsilon)=q_{\rm EA}+A\eps^{1/2} \ .
\ee
The overlap correlation function goes to zero exponentially with a 
correlation length that for $\epsilon\to 0$ diverges as 
$\eps^{-\frac14}$.  The non-integer power (less than one) in the dependence of 
$q(\epsilon)$ over $\epsilon$ implies that $\frac{dq}{d\eps} 
\big|_{\eps=0}=\infty$ and consequently the correlation length in a single phase 
is equal to infinity.  This divergence implies that the free energy is flat in 
some directions or equivalently that the system in the broken phase is always in 
a critical state.

In the same way we can add to the Hamiltonian a term proportional to 
the energy overlap, by writing

\begin{equation}
  H_{J}(\si,\tau)=H_{J}(\si)+H_{J}(\tau)- 
  \eps\mbox{$\sum^\prime$}\si_{i}\tau_{i}\si_{i'}\tau_{i'}\ .
\protect\label{EPSE}
\end{equation}
The two Hamiltonians (\ref{EPSQ}) and (\ref{EPSE}) for positive $\eps$ 
behave in a similar way, but for negative small $\eps$ at zero 
magnetic field they have a different behavior.  In the case of 
(\ref{EPSQ}) we end up with two states with negative $q$, smaller than 
$q_{\rm EA}$.  On the contrary when using the Hamiltonian (\ref{EPSE}) 
we end up with two states that have a small negative overlap.  Finding 
a discontinuity in $q$ or $q_{e}$ as function of $\epsilon$ when using 
the Hamiltonian (\ref{EPSE}) and letting $\epsilon\to 0$ is a clear 
sign of the existence of many different equilibrium states.  

\section{D=3\protect\label{S-D3}}

In this section we will discuss the crucial, physical case of $3D$ 
systems. We will start by showing how difficult these simulations 
are (because of the nature of the spin glass phase and of the 
proximity of the LCD), by discussing high $T$ simulations 
(\ref{SS-D3WARM}). We will then show that there is a phase 
transition, and that it is mean-field like (\ref{SS-D3COLD}), by 
discussing spatial correlation functions, exact sum rules, $3$ 
replica's simulations (\ref{SS-D3THRE}) and the ultrametric structure 
of the phase space. We also show that off-equilibrium dynamic 
simulations (\ref{SS-D3DYNA}) contribute to depict a very clear 
scenario.

\subsection{Statics above $T_c$\protect\label{SS-D3WARM}}

We have already explained why numerical simulations of spin glass 
systems are difficult, and why the case of $3$ dimensions is probably 
the most difficult to analyze: in the whole cold phase one has a very 
severe slowing down (and maybe even a diverging correlation length 
for all $T<T_{c}$), and the lower critical dimension is close. 

The first possible approach to this problem is to simulate the system 
in the warm phase \cite{MAPARI}: one starts from high $T$ values, 
where simulations are easy, and goes as close as possible to the point 
of phase transition (or to a $T$ point with a very high correlation 
length).  One stops where the correlation time becomes too large as 
compared to the available computer resources, or where the largest 
correlation length in the system becomes too large as compared to the 
largest system on can simulate.

We will show here some runs done on a $64\cdot 64\cdot 128$ lattice, 
with couplings $J=\pm 1$.  Here each spin is coupled with strength one 
to $26$ neighbors (in order to make the system better behaved at low 
$T$ values).  That does change non universal quantities like the value 
of the critical coupling, but does not change the universality class 
of a $3D$ system.  We always follow the (equilibrium) dynamics of two 
replicas in each realization of the couplings, and we compute their 
overlap.  For these equilibrium runs on a large lattice we have 
averaged over two realizations of the noise, and we have checked that 
sample to sample fluctuations were under control (this is natural on 
large lattices at not so low $T$ values).  We have ran from half a 
million sweeps at the higher $T$ values up to $30$ millions sweeps at 
the lower $T$ values of our runs.

\begin{figure}
  \begin{center}
    \leavevmode
    \epsffile[20 20 440 550]{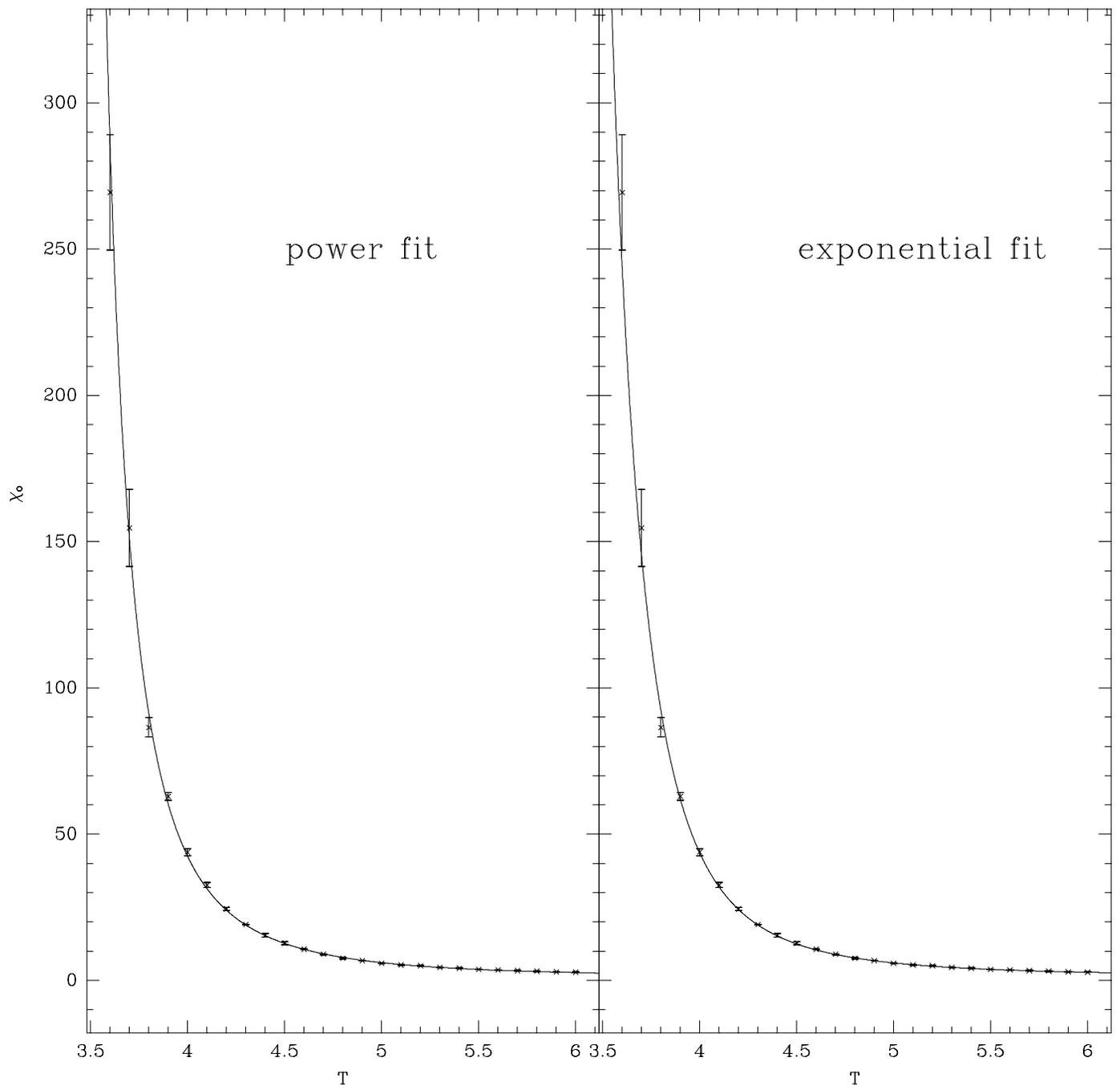}
  \end{center}
  \protect\caption[1]{The overlap susceptibility as a function of $T$, from 
  \protect\cite{MAPARI}. On the left the best power the fit, on the right 
  the best exponential fit.}
  \protect\label{F-TREOSC}
\end{figure}

In fig. (\ref{F-TREOSC}) we plot the overlap susceptibility $\chi_q$ 
as defined in eq. (\ref{E-CHIQ}). The two curves are here to give as 
the first surprise.

In the curve on the left we have tried a power fit, with a divergence 
at a finite $T_c$:

\be
  \chi_q\simeq 1+\frac{A}{\left(T-T_c\right)^\gamma}\ .
  \label{E-CHIPOW}
\ee
The best fit, in the figure, is very good, and gives $T_c=3.27\pm 
0.02$ and $\gamma=2.43\pm 0.05$. One can be happy, and believe she 
has exhibited the correct critical behavior, till then another 
functional form is tried. We have tried the $T=0$, exponential 
divergence

\be
  \chi_q\simeq A
  \left(e^{\left(\frac{B}{T}\right)^p}-1
  \right)+C\ ,
\ee
that is a very natural behavior if we are at the lower critical 
dimension. The fit is in the curve on the right, and it is again very 
good. So, we find that a fit that looks very good does not give much 
information about the nature of the critical region.

We have also considered the correlation length defined in 
(\ref{E-CSISI}).  Also here a power fit to a divergence at a finite 
$T_c$ works very well, giving a value of $T_c$ compatible with the one 
we have seen before, and an exponent $\nu=1.20\pm 0.04$.  Also in this 
case the exponential fit works very well (even better than the power 
fit), and gives for the parameters values that are consistent with the 
ones we found for $\chi_q$.  It is also interesting to note that we 
have tried a large number of fits, that all give a fair description of 
the behavior of the system in the critical or transient region.  For 
example a fit of $\chi_q$ to the form $ \exp(A\exp(B\beta))$
also works very well.

So, the problem is difficult.  Since the lower critical dimension is 
close (maybe at zero distance) it is difficult to be sure that we are 
really dealing with a finite $T$ divergence.  We will see that in 
order to be sure of the existence of a phase transition one has to be 
able to go deep in the cold region on large lattices \footnote{The 
finite size scaling analysis of small lattices leads to ambiguities 
very similar to the ones we have described here.}, and that using the 
tempered Monte Carlo makes this goal far easier.

\subsection{Statics  at $T_c$ and below $T_c$\protect\label{SS-D3COLD}}

The first results that have recently made clear the existence of a 
phase transition are the ones obtained by Kawashima and Young 
\cite{KAWYOU}.  We will discuss these and the recent unpublished 
results by Marinari, Parisi and Ruiz-Lorenzo \cite{MPRUNP}.  Only 
after that we will discuss the characterization of the cold phase 
\cite{MPRR}, by ignoring the temporal sequence of the papers (it turns 
out that by analyzing correlation functions and observables related to 
the $P(q)$ it is easier to characterize the regime of low $T$ as a 
mean-field like regime than to be sure that there is a real phase 
transition and not only a $T=0$ exponential divergence of the 
correlation length in the overlap sector of the theory).

\begin{figure}
  \begin{center}
    \leavevmode
    \epsfysize=250pt\epsffile{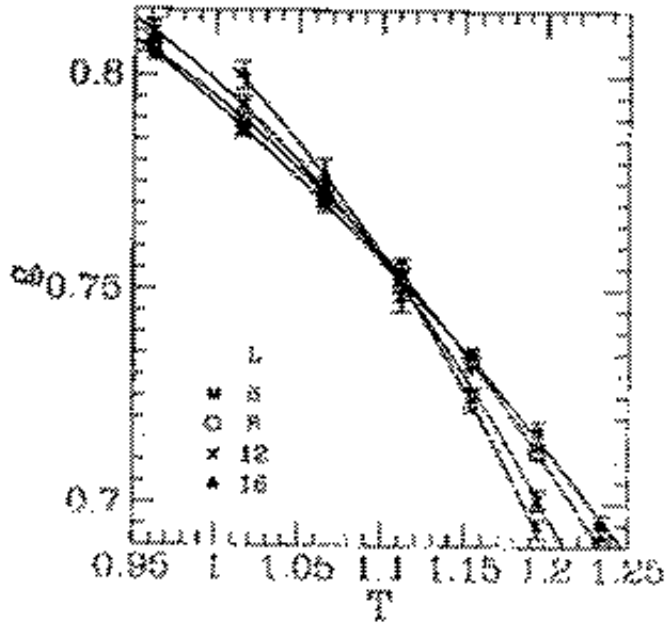}
  \end{center}
  \caption[0]{Binder cumulant for the $3D$, $J=\pm 1$ spin 
   glass. From \protect\cite{KAWYOU}.}
  \protect\label{F-KAWYOU-1}
\end{figure}

Kawashima and Young \cite{KAWYOU} have studied a $3D$ spin glass on a 
simple cubic lattice, with coupling $J=\pm 1$.  They are able 
to thermalize under $T_c$ lattices of size going up to $16^3$.  They 
use a large number of samples (from $8000$ to $2000$ for the different 
lattice sizes), with a number of sweeps going from $.5$ to $15$ 
millions: nine equivalent years of IBM $390$ processor, a good show of 
a brute force approach.  In figure (\ref{F-KAWYOU-1}) we show their 
Binder parameter $g$ (\ref{E-BINDER}) in the critical region.  At 
$T=1.0$ they can exhibit a statistically significant crossing of the 
Binder parameter: it is a small effect, but now significant at a few 
standard deviations (two or three).  It is interesting to notice that 
the lower $T$ value where they can get the $16^3$ lattice to thermal 
equilibrium is $T^{(\min)}\simeq 0.9 T_c$: it is very difficult to 
thermalize at low $T$ values, and we will see that tempering is 
crucial for that.  

\begin{figure}
  \begin{center}
    \leavevmode
    \epsfysize=250pt\epsffile{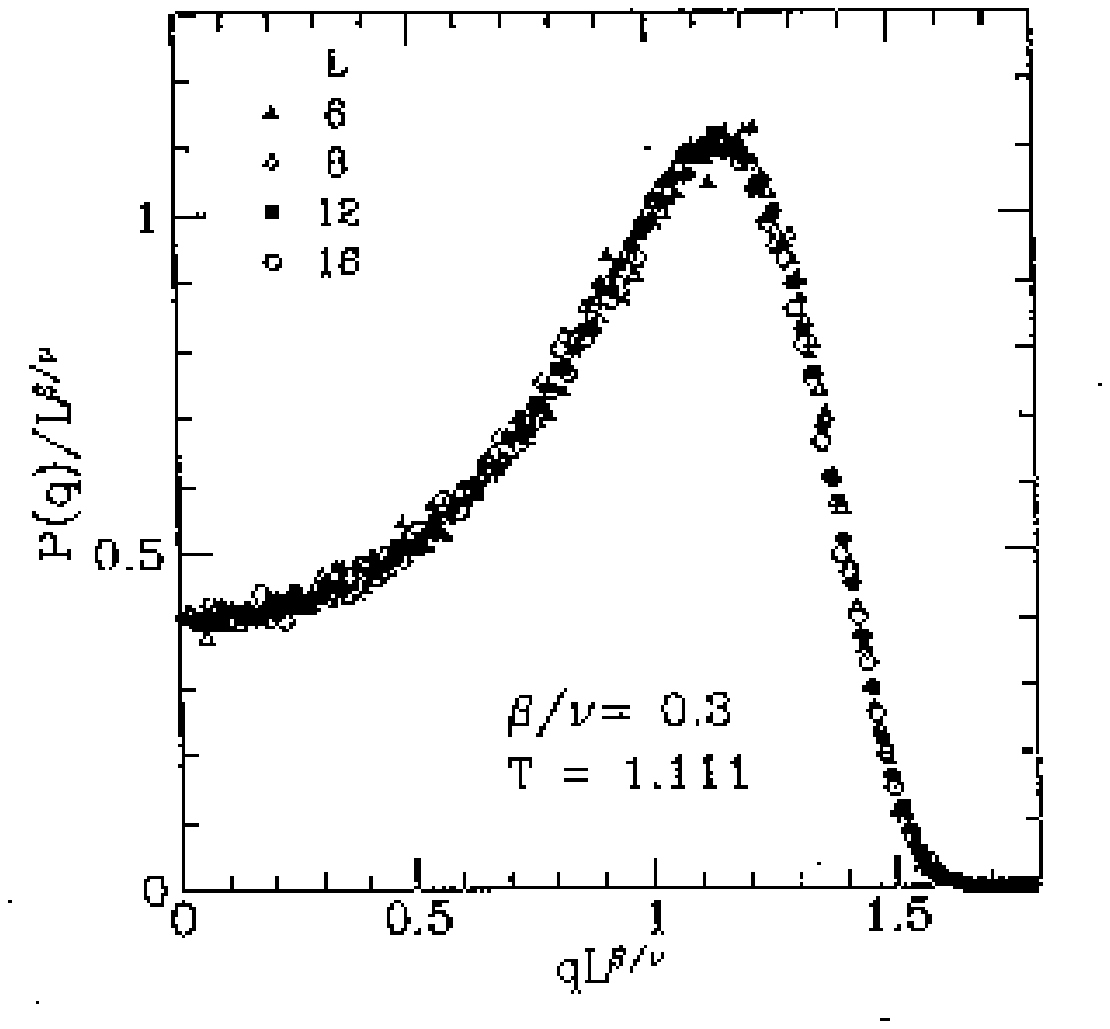}
  \end{center}
  \caption[0]{Rescaled $P(q)$ near the critical point for the $3D$ $J=\pm 
  1$ spin glass.  From \protect\cite{KAWYOU}.}
  \protect\label{F-KAWYOU-2}
\end{figure}

One can also use the probability distribution of the overlap, $P(q)$, 
computed at the critical point, to determine critical exponents.  One 
uses the relation

\be
  P(q)=L^{\frac{\beta}{\nu}} 
  f\left(q L^{\beta/\nu},L^{\frac{1}{\nu}}(T-T_c)\right)\ ,
\ee
for $T=T_c$ to estimate the ratio $\frac{\beta}{\nu}$.  We show the 
result of the best fit in figure (\ref{F-KAWYOU-2}): one finds 
$\frac{\beta}{\nu} \simeq 0.3$.  The best determination of Kawashima 
and Young \cite{KAWYOU} of the critical value of $T$ and of the 
critical exponents is $T_c^{(\pm 1)}=1.11\pm 0.04$, $\nu=1.7\pm 0.3$ 
and $\eta = -0.35 \pm 0.05$.

\begin{figure}
  \begin{center}
    \leavevmode
    \epsfysize=250pt\epsffile{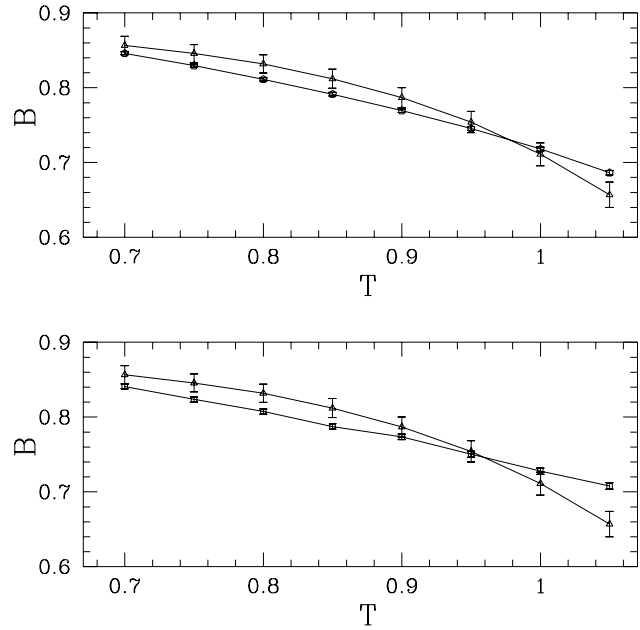}
  \end{center}
  \caption[0]{Binder cumulant for the $3D$ spin glass with Gaussian 
  couplings \cite{MPRUNP}. In the lower plot $L=4$ and $L=16$, in the 
  upper one $L=8$ and $L=16$.
  }
  \protect\label{F-MPRBIN}
\end{figure}

Recent results have been obtained in \cite{MPRUNP} by using the 
parallel tempering Monte Carlo (\ref{S-PARTEM}) to simulate the $3D$ 
EA first neighbor spin glass model with Gaussian couplings \footnote{ 
These simulations have been ran on the APE parallel supercomputer 
\cite{APE93}.}.  The use of an improved Monte Carlo technique has 
allowed to thermalize lattices of size up to $16$ down to 
$T^{(\min)}\simeq 0.7 T_c$ (a large gain over what was possible with 
the standard Monte Carlo approach).  In fig.  (\ref{F-MPRBIN}) we plot 
the Binder parameter for $L=4$ and $L=16$ (lower plot), and for $L=8$ 
and $L=16$ (upper plot).  In both cases the crossing is statistically 
significant in a whole set of $T$ values.  It is also interesting to 
look at the value of the Binder parameter at the critical point, that 
is an universal quantity: the two cases of $J \pm 1$ and of Gaussian 
couplings give compatible values, close to $0.75$.  This fact 
constitutes one more evidence for the existence of a phase transition 
in $3D$.

After establishing the existence of a phase transition in $3D$, we 
will clarify (mainly after the simulations of \cite{MPRR}) the nature 
of the cold phase.  Again, we are weighting here two possible 
behaviors: the predictions of Mean Field theory with spontaneous 
replica symmetry breaking (e.g.  a large number of pure states) and 
the ones from the droplet model (e.g.  only two pure states).  In 
order to try and solve this issue we will discuss here about two main 
sets of observables: {\em i}) the behavior of the overlap-overlap 
correlation function when the overlap is close to zero as a function 
of space and time; {\em ii}) the behavior of the Binder cumulant 
computed on blocks of different sizes as a function of the block size 
and of the Monte Carlo time.

Since it is practically impossible to equilibrate very large lattices 
at very low $T$ values, a shortcut can help: one can for example 
analyze the dynamic behavior of the system to get information about 
the equilibrium structure.  That is why we are discussing the results 
of \cite{MPRR} in this section and not in the section about dynamics: 
here one uses a dynamic behavior together with an ansatz on the rate 
of the convergence to equilibrium to get equilibrium information.  In 
the section about off-equilibrium simulations we will describe 
numerical experiments where one is dealing with quantities that 
represent intrinsically off-equilibrium phenomena: here, on the 
contrary, we use off-equilibrium dynamics and a reasonable and 
verified guess about convergence to equilibrium in order to derive 
properties of the thermalized system.

So, one \cite{MPRR} simulates large lattices {\em to avoid} the 
equilibrium situation: starting from random initial conditions one 
gets a value of the overlap close to zero, that stays close to zero 
during all the MC run (one needs a huge number of MC sweeps to 
start and form a macroscopic overlap on a very large lattice 
\cite{MPRR}).

Let us consider two copies of an infinite system.  In practice one 
takes a system whose size is much larger than 
$t_{\max}^{{1}/{z(T)}}$, where $z(T)$ is the appropriate dynamic exponent.  
The overlap $q$ among the two copies at $t=0$ is zero, since one 
selects two random configurations, and it remains close to zero during 
all the $t_{\max}$ MC sweeps.  In this way the local correlation 
functions go to a finite limit and they are interpreted to be those of 
two equilibrium states at $q=0$.  It is trivial to verify that in the 
case of a ferromagnet (or more generally of a system with a unique 
equilibrium state, neglecting reflections) one finds that

\be
  G_x \rightarrow q_{\rm EA}^2\;\;\;{\rm as} \;\;\; x\rightarrow 
  \infty\ ,
\ee
where $G_x$ has been defined in (\ref{E-GJ}).

At time $t_0$ one quenches the system to $T<T_c$, and starts measuring 
the overlap-overlap correlation function $G_x(t)$ of eq.  (\ref{E-GJ}) 
(computed now only at time $t$) at distance $x$ and time $t$.  At a 
given time $t$ the system is correlated up to a distance of the order 
of the dynamic correlation length $\xi(T,t)$, i.e. the correlation 
functions are statistically different from zero up to this distance. 
The dynamic correlation length $\xi(T,t)$ grows in time as

\be
  \xi(T,t) \propto  t^{\frac{1}{z(T)}}\ ,
\ee
that defines the dynamic critical exponent, $z(T)$ (in the pure 
Ising model at $T_c$ $z=2$, while in the SK model $z(T_c)=4$).  In 
this way we are trying to verify a power law increase of the dynamic 
correlation length in all the broken phase, for $T<T_c$.  $z(T)$ can 
(and does) depend on the temperature $T$.

The numerical data follow very well the functional form

\be
  G_{x}(t) = \frac{A(T)}{x^\alpha} \exp \left\{-\left( \frac{x}{\xi(T,t)}
  \right)^\delta  \right\} \ ,
  \label{cor:g}
\ee
in a wide rage distance and time regions.  Numerical data support this 
behavior in all the region that has been analyzed, i.e.  for $1\le x 
\le 8$, $10^2 \le t \le 10^6$ and $0.3~ T_c \le T \le T_c$.  In all 
these simulations the value of the overlap, $q$, remains very close to 
zero (since the lattice is large enough as compared to the observation 
time).  One finds that \cite{MPRR} $z(T)\simeq \frac{6.25}{T}$, an 
estimate compatible with the results of \cite{KISASCHRI96}.  For 
example one estimates $z(T_c)=6.25\pm 0.30$, in good agreement with 
the results of \cite{OGIELSKI85}($z(T_c)=6.1\pm 0.3$), the ones of 
\cite{BLUHUBRAY92} ($z(T_c)=5.85\pm 0.30$) and the ones of 
\cite{BHAYO92} ($z(T_c)=6.0\pm 0.5$).  The exponents $\alpha$ and 
$\delta$ show very little dependence on $T$: for example at $T=0.70$ 
one finds $\alpha=0.50\pm 0.02$ and $\delta=1.48\pm 0.02$.

\begin{figure}
  \begin{center}
    \leavevmode
    \epsfysize=250pt\epsfxsize=300pt\epsffile{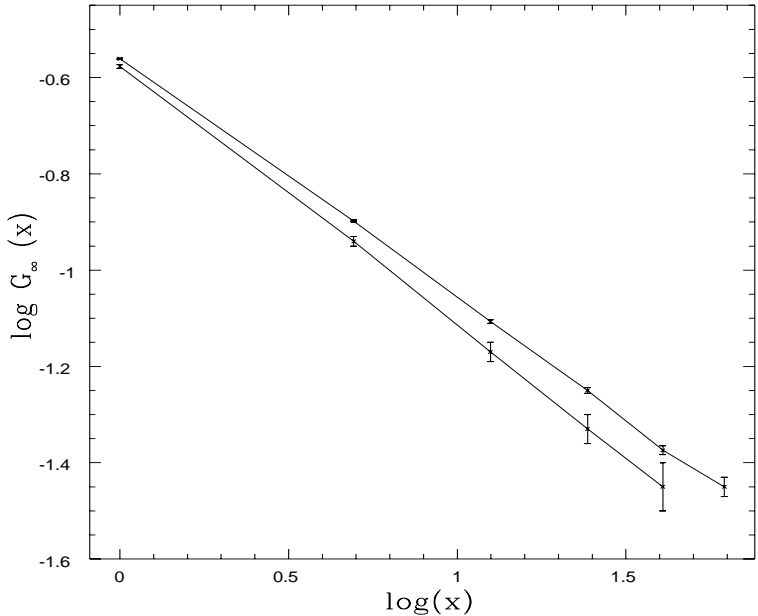}
  \end{center}
  \protect\caption{$G_\infty(x)$ against $x$ in a dilogarithm scale 
    ($T=0.7$). The
    upper line is the result of a slow cooling while the lower one
    is obtained after a sudden quench to $T<T_c$ (see text).}
  \protect\label{fig:cor_q0}
\end{figure}

An effective way to proceed is to take the $t\to\infty$ limit at 
fixed $x$ on the numerical data, by using the form (\ref{cor:g}).
This procedure gives consistent results, and one obtains in this way 
data that will be fitted as

\be
  G_x(t=\infty) \equiv \lim_{t \rightarrow \infty } G_x(t) 
  =\frac{A(T)}{x^\alpha}\ .
  \protect\label{res_g}
\ee
We show this correlation function in figure (\ref{fig:cor_q0}) together
with the extrapolated correlation function obtained using a cooling
procedure. The power law behavior is very clear.

In the mean field framework it is possible to get analytic predictions 
for these decays. de Dominicis and Kondor \cite{DOKO94} have used RSB 
theory to compute the $q-q$ correlation function restricted to the 
$q=0$ sector of the phase space, $G_x^{(q=0)}$.  One expects a power 
law behavior, i.e.

\be 
  G_x^{(q=0)} \simeq \frac{1}{|x|^{\hat \alpha}}\ .
  \protect\label{dominicis}
\ee
So, there is a good agreement of the expectation generated by the mean 
field picture and the numerical results: correlation functions in the 
$q=0$ sector have an equilibrium limit and decay like a power law.  
These features could not be explained by a droplet model like picture, 
where there are no $q=0$ equilibrium correlation functions, and the 
only correlation functions of the theory eventually have to decay to a 
constant (the square of the EA order parameter, $\qea^2$).  This 
evidence is strongly favoring a mean field like picture.

What we have been discussing in the last paragraphs concerns ergodic 
components of the phase space.  We have shown that correlations in the 
$q=0$ ergodic component of the $3D$ system can be measured, and that 
one can detect a power law decay, that is what one expects from the 
mean field theory.  We will see now that one can get even stronger 
evidence that the stable states of the system are organized in a non 
trivial structure.  Thanks to a sum rule we will be able to compare 
\cite{MPRR} the full correlation at distance $1$ and the correlation 
in the $q=0$ sector, and we will show that they are different in the 
broken phase, for $T<T_c$.

In the case of Gaussian couplings, by integrating by parts the 
expression for the expectation value of the link energy operator it is 
easy to obtain

\be
   E_{\rm link}=-\beta(1-G_{x=1}) \ ,
\ee
that relates the expectation value of the energy (that can be 
determined with high precision from the numerical data) to the 
correlation function (integrated over all ergodic components,
(\ref{E-GALL})) at a distance of one lattice spacing.

 The value of energy is well 
determined in the numerical simulation.  One can extrapolate to infinite time by 
using the form $E(t)=E_\infty +A t^{-\Delta(T)}$.  The fit works well: the 
exponent $\Delta(T)$ is reasonably large.  One estimates \cite{MPRR} that 
$\Delta(T) = 0.44 T = \frac{2.75}{z(T)}$.  This compares very well a mean field 
computation \cite{FRPAVA,FRPAVB} based on the analysis of the interface free 
energy, where one finds $\Delta(T) \frac{2.5}{z(T)}$: one more quantitative 
prediction of the mean field theory that describes very well the $3D$ case.  One 
gets a good estimate for $E_\infty$.  This in turn gives a precise estimate of 
$G_{x=1}$ One finds that in the high $T$, paramagnetic phase, the $q=0$ 
correlation functions equals, as expected, the full function, i.e.

\be
  G_{x=1}^{(q=0)} = G_{x=1}\ ,
  \protect\label{ergo}
\ee
where as we have explained we have identified the correlation 
function measured at short times with the $q=0$ average, and the 
equality works in the warm phase with a precision better than one 
percent.  On the contrary as soon as we enter in the cold phase the 
equality (\ref{ergo}) is violated: for example at $T=0.7$ one has 
$G_{x=1}^{(q=0)}=0.612\pm 0.001$ and $G_{x=1}=0.56\pm 0.01$ while at 
$T=0.35$ $G_{x=1}^{(q=0)}=0.802\pm 0.001$ and $G_{x=1}=0.67\pm 0.01$.  
This is a strong indication that there are more ergodic components, 
i.e.  that the replica symmetry is broken.

\begin{figure}
  \begin{center}
    \leavevmode
    \epsfysize=200pt\epsffile{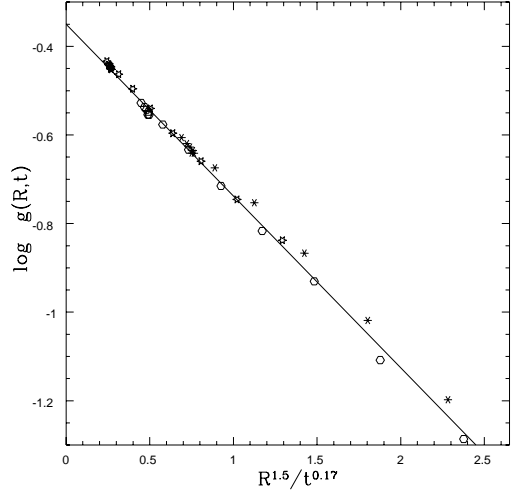}
  \end{center}
  \protect\caption{The logarithm of the Binder cumulant for the box overlap
    versus rescaled ratio of time and distance. Stars are for $R=2$,
    hexagons are for $R=3$ and asterisk for $R=4$. The straight line is
    only a guide the eye.}
  \protect\label{fig:binder}
\end{figure}

We will describe now a last numerical experiment that is also meant to 
detect a difference of the DM scenario and the RSB mean field 
approach.  We will see that again a DM approach is falsified from the 
numerical findings.  The experiment is based on studying the quantity 
$P(q_R)$, i.e.  the probability distribution of the overlap in a box 
of linear size $R$, $q_R$.

In the RSB solution of the Mean Field theory the probability 
distribution $P(q_R)$ is Gaussian for $R\rightarrow \infty$, 
$\frac{R}{L}\ll 1$, while in a DM inspired solution it converges to 
the sum of two Dirac delta functions (one in $+q_{\rm EA}$ and another 
in $-q_{\rm EA}$). A practical way to discern among the two 
possibilities is to look at the Binder cumulant. At time $t$ the 
cumulant for block of size $R$ is defined as

\be
  g(R,t)\equiv 
  \frac{1}{2}\left(3-\frac{\overl{\langle q_R^4 \rangle }}
  {\overl{ \langle q_R^2 \rangle}^2}\right)\ ,
\ee
where $g(R,t)$ is built on data measured after $t$ MC sweeps.
>From standard dynamic scaling one expects

\be
  g(R,t)=f(\frac{R}{\xi(t)})\ ,
\ee
where $f$ is a scaling function.  In figure (\ref{fig:binder}) we show 
the data for $T=0.7$ and $R=2$, $3$ and $4$.  We plot the logarithm of 
the block Binder cumulant versus 
$\left(\frac{R}{t^{\frac{1}{z}}}\right)^{\delta}$, by using the 
exponents $\delta$ and $z$ determined before from the behavior of the 
overlap-overlap correlation functions (i.e.  $\delta=1.5$ and 
$z=8.3$). The figure makes clear we are not dealing with a $\delta$ 
function (that would be characterized by $\log(g)=0$): analyzing the 
system on larger and larger scales we do not find a ferromagnetic 
behavior, disproving again a droplet like picture.

\subsection{Simulations with Three Replicas\protect\label{SS-D3THRE}}

One of the potential advantages of using three replicas (i.e. $3$ 
copies of the system with the same quenched couplings $J$) in a 
numerical simulation is the possibility of investigating more details 
of the $P(q)$ (for example by defining new, different Binder cumulant 
like parameters, and trying to understand if they exhibit a clearer 
critical behavior: critical exponents are universal, but amplitudes 
are not). Also, as we will discuss in some detail, working on $3$ 
replicas helps in getting hints about the metric (or ultrametric) 
structure of the phase space \cite{RATOVI,CAMAPA}.

Here we will introduce a Binder cumulant that allows to observe a 
crossing of curves, plotted as a function of $T$, obtained for 
different lattice sizes $L$: that strengthens the results about the 
existence of a phase transition that we have already discussed.  In 
the following (\ref{SS-D4ULTR}) we will discuss a detailed study of 
ultrametricity in the $4D$ model done by using a similar approach.
Here we are discussing about equilibrium simulations.

Let $\sigma, \tau$ and $\mu$ be three replicas of our $3D$ spin glass: 
we will simulate them in parallel, using the same quenched disorder 
(and different random numbers for the dynamic).  We will define three 
different overlaps that we will denote $\{q_{12}, q_{23}, q_{13}\}$ or 
$\{q, q^\prime, q^{\prime \prime} \}$ in the rest of this subsection 
(where we will mainly follow \cite{INPARU96}).

In (\ref{E-GUERRA}) we have shown one typical relation among 
expectation values of the probability distribution of the overlap.  
These relations embody the ultrametric content of the mean field 
theory \cite{MEPAVI,BREAKS}. Two specific cases can be written as

\be
  \overline{ \lan q^2 \ran ^2 }
  = \frac{1}{3} \overline{ \lan q^4 \ran} +
    \frac{2}{3} \overline{ \lan q^2 \ran}^2 \ ,
  \protect\label{FOR_1}
\ee

\be
  \overline{ \lan q^2 {q^{\prime}}^2 \ran}
  =  \frac{1}{2}\overline{\lan q^4 \ran} + 
     \frac{1}{2}\overline{\lan q^2 \ran}^2\ .
  \protect\label{FOR_2}
\ee
Recently Guerra \cite{GUERRA95} succeeded to obtain some of these in a 
rigorous approach to spin glass theory, proving the validity of a set 
of such relations even for finite dimensional models (constructed by 
sending to zero a mean-field like perturbation of the Hamiltonian): 
these results justified the numerical findings of \cite{MPRR}.  Both 
(\ref{FOR_1}) and (\ref{FOR_2}) have been analyzed in detail in 
\cite{INPARU96}: one finds small finite size corrections, and a very 
satisfactory agreement of the numerical data and the theoretical 
result in the infinite volume limit.  After Guerra \cite{GUERRA95} 
results establishing the numerical validity of (\ref{FOR_1}) and 
(\ref{FOR_2}) can be considered as a good check of the thermalization 
(and of the formal correctness of the computer codes!).  

\begin{figure}
  \begin{center}
    \leavevmode
    \epsfysize=250pt\epsffile[47 79 447 633]{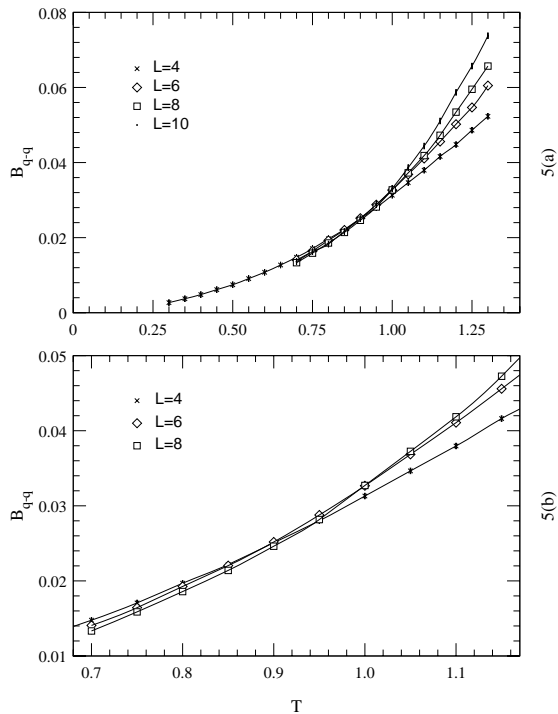}
  \end{center}
  \protect\caption{ In the upper figure
  $B_{q-q}$ versus $T$ for $L=4$ to $10$ in a large
  $T$ range. In the lower one $L=4,6,8$ and $T\in 
  [0.7,1.3]$.}
  \protect\label{fig:3d3}
\end{figure}

As we said by running simulations of $3$ copies of the system one can 
define more cumulants, that can allow to extract more information 
about the system. Following \cite{INPARU96} one defines

\be
  B_{qqq} \equiv 
  \frac{\overline{\lan \vert q_{12}q_{13}q_{23} \vert 
  \ran}}
  {\overline{\lan q^2 \ran}^{3/2}}\ , \ \ \ \
  {B^{\prime}}_{qqq} \equiv \frac{\overline{\lan q_{12}q_{13}q_{23} 
  \ran}}
 {\overline{\lan q^2 \ran}^{3/2}}\ ,
  \protect\label{FOR_3}
\ee
and

\be
  B_{q-q} \equiv \frac{\overline{\lan {(\vert q_{12} \vert -
  \vert q_{13} \vert)}^2 \ran}}
  {\overline{\lan {q_{23}}^2 \ran}}\ , \ \ \ \ \
  {B^{\prime}}_{q-q} \equiv \frac{\overline{\lan {( q_{12} -
  q_{13}\sign(q_{23}))}^2 \ran}}
 {\overline{\lan {q_{23}}^2 \ran}}\ ,
  \protect\label{FOR_4}
\ee
where $q_{23}$ is the largest of the three overlaps (in absolute value). 
One expects that standard finite size scaling applies:

\be
  B_{\#} = f_{\#}(L^{\frac{1}{\nu}} (T-T_c))\ ,
\ee
where we have used the symbol $\#$ to denote one of the cumulants we 
have just defined.  $B_{qqq}$ and $B^{\prime}_{qqq}$ turn out to have 
the same behavior than the usual Binder cumulant based on two replicas 
(see figures (\ref{F-KAWYOU-1}) and (\ref{F-MPRBIN})).  $ B_{q-q}$
seems instead to show a clearer signature of the phase transition: in 
figure (\ref{fig:3d3}) we show the $L=4$, $6$ and $8$ data.

Let us finally quote some preliminary results about the ultrametric 
structure of the phase space of the $3D$ model \cite{INPARU96}.  One 
starts by measuring, after each MC iteration, the $3$ overlap among the 
$3$ copies of the system, and ordering them in $q_{\max}$, 
$q_{\mbox{med}}$ and $q_{\min}$. One defines

\be
  b \equiv \frac
  { {\left (\vert q_{\mbox{med}} \vert - \vert q_{\min} \vert\right)}^2 }
  { {q_{\max}}^2 }\ .
\ee
One defines the integrated probability $\Pi(b>b_0)$ by

\be
\Pi(b>b_0)\equiv\int_0^{b_0} {\rm d}b~ P(b)\ .
\ee
In the small $b_0$ region one finds that $\Pi(b_0)$ decays with a 
power law, i.e.  $\Pi(b_0)\simeq b_0^{-\alpha}$.  For intermediate 
values of $b_0$ one sees a fast, exponential decay $\Pi(b_0)\simeq 
e^{-\beta b_0}$, while in the large $b_0$ region $\Pi(b_0)$ goes to 
zero faster than an exponential.  One can also fix $b_0$ (for example 
by taking $b_0=0.05$): $\Pi(b_0=0.05)$ decays as power law with the 
size of the system $L$.  In an ultrametric phase space $P(b)$ is a 
$\delta$ function centered in the origin: these results suggest that 
ultrametricity holds in $3D$.  Also, the theoretical analysis of 
\cite{INPARU96}, based on the results of \cite{GUERRA95}, shows that 
if the phase space of a finite dimensional system is ultrametric than 
necessarily equations like (\ref{E-GUERRA},\ref{FOR_1}-\ref{FOR_4}) 
must hold, i.e.  one must find the same ultrametric structure of the 
mean field solution.

\subsection{Out of Equilibrium Dynamics\protect\label{SS-D3DYNA}}

In the following we will discuss about out of equilibrium dynamics of 
the $3D$ EA spin glasses.  We do not have enough space to give more 
than basic information: we will be mainly discussing the work by 
Rieger and coworkers \cite{KISASCHRI96,RIEGER93,RIEGER94,RIEGER96}, 
that the interested reader should consult.  The crucial points can be 
summarized in a few words.  In first numerical simulations give 
results that are completely compatible with the experimental results 
(concerning, for example, the decay of magnetization after switching 
off an applied field).  Aging phenomena \cite{BOU} are clear.  In 
second the most part of results are not compatible with the 
logarithmic dependence on time implied by the droplet picture.  Aging 
phenomena turn out to be clearly characterized by functions 
$f(\frac{t}{t_w})$ and not, as the droplet model would imply, by 
functions of $\log(\frac{t}{\tau}) / \log(\frac{t_w}{\tau})$.

\begin{figure}
  \begin{center}
    \leavevmode
    \epsfxsize=250pt\epsffile[38 230 400 520]{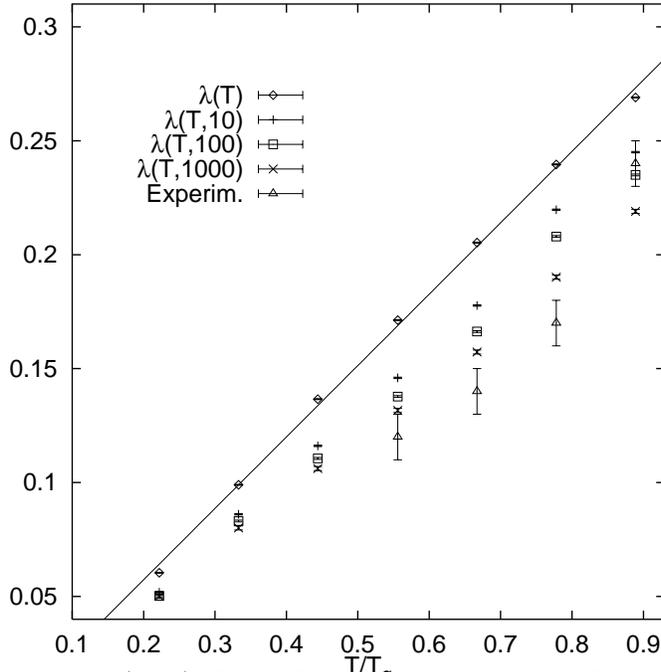}
  \end{center}
  \protect\caption[0]{The non-equilibrium 
    exponent $\protect{\lambda(T,t_w)}$ of the $3D$
    EA-model. The straight line is a linear 
    fit of $\protect{\lambda(T)}$ and
    is a guideline for the eye only. 
    From \protect\cite{KISASCHRI96}.}
  \protect\label{fig:lambda}
\end{figure}

One measures autocorrelation functions at different times, and tries 
to determine the functional form of the power decay: we will see that 
numerical results can be well compared to real experimental results.
The remnant magnetization, measured at time $t$ after a sudden quench 
(when a large applied magnetic field is switched off), is defined as

\begin{equation}
  M(t) \equiv C(t,0)\ .
\end{equation}
Experiments show a clear power law decay, i.e.

\begin{equation}
  M(t) \sim t^{-\lambda(T)}\ ,
  \protect\label{tmr}
\end{equation}
where $\lambda(T)$ depends on the temperature.  In figure 
(\ref{fig:lambda}) we show $\lambda$ versus $T$ (from 
\cite{KISASCHRI96}: the experimental exponents are from \cite{MANO}).  
In figure (\ref{fig:lambda}) are also the exponents $\lambda(T,t_w)$, 
obtained by looking at the decay of $C(t,t_w)$, for values of the 
waiting time $t_w \ll t$.  The data from the real experiment are from 
the remnant magnetization measurements in an amorphous metallic spin 
glass.  Even if there is a quantitative difference among the numerical 
and the experimental values the data are very similar 
(we are discussing about critical exponents, that are always measured 
with a quite high uncertainty, often more of a systematic than 
statistical nature).

The autocorrelation function $C(t,t_{w})$ (\ref{E-CORDYN}) can be 
analyzed in two different regimes:

\begin{itemize}

\item The fully off equilibrium regime (where there is no invariance 
under time translation),  $t \gg t_w$ \cite{FRARIE95}. The 
asymptotic decay of the remnant magnetization that we have discussed 
before is a special case ($t_w=0$).

\item The {\it quasi} equilibrium regime that the system 
reaches for $t\ll t_w$.

\end{itemize}

The behavior of the autocorrelation function in the two cases is well 
described as

\begin{equation}
  \protect\label{corre3d}
  C(t,t_w) \sim 
  \left\{ 
    \begin{array}{ll}
       t^{-\lambda(T,t_w)} & \mbox{ if }  t \gg t_w\ ,\\
       t^{-x(T)}           & \mbox{ if }  t \ll t_w\ .
    \end{array}
  \right.
\end{equation}
The droplet model would imply in the region $t \gg t_w$ a behavior 
$C(t,t_w) \sim (\log t)^{-\lambda/\psi}$ that does not describe well 
the numerical data.

We can write (\ref{corre3d}) in a compact form by defining a scaling 
function $f$ such that $C(t,t_w)=t^{-x} f(t/t_w)$.  The scaling 
function $f(z)$, tends to a constant when $z \rightarrow 0$ and 
behaves as $z^{-\lambda +x}$ as $z \rightarrow \infty$: $\lambda$ and 
$x$ are the exponents defined in equation (\ref{corre3d}).

The prediction of the droplet model for the correlation function is

\begin{equation}
  C(t,t_w)=(\log t)^{\theta/\psi} g\left(\frac{\log(t/\tau)}
  {\log(t_w/\tau)}\right)\  ,
\end{equation}
where $\theta$ and $\psi$ are the droplet model exponents and $\tau$
is a time scale. Also this fit turns out to be inadequate to describe 
the numerical data. The naive droplet model is definitely falsified 
from the off-equilibrium dynamic simulations (and from the 
experimental data). On the contrary mean field theory is 
characterized by power law decays.

It is also interesting to study the domain growth.  One looks at the 
autocorrelation function among overlaps (see equation (\ref{E-GJ})).  
One defines a dynamic correlation length as

\be
  \xi(t_w)=2 \int_0^\infty {\rm d} r G_r(t_w)\  .
\ee
The dynamic correlation length, $\xi(t_w)$, turns out to be described 
very well by an algebraic behavior $\xi(t_w)\sim t_w^{\alpha(T)}$, 
where the exponent $\alpha$ depends linearly over $T$.  In this case 
also the droplet model behavior $\xi(t_w)\sim (\log t_w)^{1/\psi}$ 
fits the data, with $\psi=0.71\pm 0.02$.

Another interesting result has been obtained in \cite{FRARIE95}.  One 
computes the ratio between the response ($R(t,t^\prime)$) and the time 
derivative of the autocorrelation function $C(t,t^\prime)$.  If the 
fluctuation-dissipation theorem holds this ratio must be equal to the 
inverse temperature $\beta$, but in the general case of a complex 
off-equilibrium dynamics we expect \cite{CUKU} that

\be
  \beta \ x_{\rm d}(t,t^\prime)= \frac{ R(t,t^\prime)}{
  \displaystyle{\frac{\partial C(t,t^\prime)}{\partial t^\prime}}}\ .
\ee
On general grounds one can expect that the a priori arbitrary function 
$x_{\rm d}(t,t^\prime)$ would in reality only depend on $C(t,t^\prime)$ which 
is the dynamic equivalent of the overlap $q$.  In this case $x_{\rm d}(q)$ can 
be interpreted as the off-equilibrium version of the function $x(q)$ of the 
static case \cite{BREAKS}.  Since at equilibrium $q\to\qea$ one recovers the 
fluctuation-dissipation theorem (since $x(q_{\rm EA})=1$).  

\section{D=4\protect\label{S-D4}}

The $4D$ case is somehow easier to study numerically than the $3D$
model. The evidence for the existence of a broken phase with a non 
trivial $P(q)$ and of a mean field like behavior is easy to achieve. 
Because of that the $4D$ model will be discussed here from two points 
of view. In first it will be seen as the model 
where firm evidence for the mean field pattern to apply in finite 
number of dimensions has been established. In second it will be 
discussed as the model where more difficult questions, like the 
existence of an ultrametric organization of the phase space, start to 
be analyzed in detail.

\subsection{Close to the Phase Transition\protect\label{SS-D4NEAR}}

First Bhatt and Young \cite{BHAYO85,BHAYO88,REBHYO} noticed that in 
the $4D$ EA model one can locate $T_c$ with a relatively small amount 
of computational work.

In $4D$ the curves representing the overlap Binder cumulant as a 
function of $T$, for different size values $L$, cross very clearly 
giving a precise estimate of $T_c$ (as a function of increasing 
lattice size the cumulant tends to zero from above in the warm phase, 
and to a non-trivial, non-zero value from below in the broken phase: 
the $T$ point where different $L$ curves cross is a good finite size 
estimate of the infinite volume $T_c$).  The very clear crossing (a 
behavior similar to the one seen in the SK model or, for the 
magnetization cumulant, in the $3D$ Ising model) allows a precise 
estimate of $T_c$ ($T_c=2.02\pm 0.03$ for $J=\pm 1$, $T_c=1.75\pm 
0.05$ for Gaussian couplings: see \cite{BHAYO85,BHAYO88,REBHYO} and 
the more recent simulations of \cite{BADONI}, done using the dedicated 
parallel computer RTN \cite{RTN}).

The value of the critical exponent $\nu$ turns out to be quite small 
(about $0.8$).  This value is nearly a factor $2$ smaller that the 
three dimensional value and this implies that on a finite lattice we 
can go much closer to the critical point by keeping finite size 
effects small.

\subsection{Below the Transition\protect\label{SS-D4UNDE}}
 
The most interesting results have been obtained by simulations done 
below the phase transition point.  The measurements of the overlap 
probability distribution $P(q)$ can be done at $T<T_c$ much easily 
than in the three dimensional case.  Finite size effects turn out to 
be large (this is already true in the SK model, and stays true three 
dimensions: it looks like an intrinsic problem of systems with 
quenched disorder).  The variation of $P(q)$ as function of the 
lattice size (for $L$ going from 3 to 7) is of the same order of 
magnitude than the one one finds in three dimensions) 
\cite{PARI93,CIPARI93}.

Thermalization is faster here than in $3D$ and good quality results in 
a large region of the broken phase have been obtained by using simple 
minded Monte Carlo techniques.  The probability distribution of the 
energy overlap converges in the infinite volume limit to a non trivial 
function.

The difference in the crossing properties of the Binder parameter in four and in 
three dimensions has clear origin.  If replica symmetry is spontaneously broken, 
in the infinite volume limit the Binder parameter converges to a non trivial 
function of $T$, $g(T)$.  In the mean field theory \cite {BREAKS}the function 
$g(T)$ for $T<T_c$ is approximately given by $1- .4{\theta(1-\theta)}$ (we have 
defined by $\theta$ the reduced temperature, $\theta \equiv \frac{T}{T_c}$).  In 
other words $g_{-}\equiv \lim_{T \to T_{c}^{-}}g(T)$ is $1$, which is quite 
different from the value of the Binder parameter {\em at} the crossing point 
(which is close to .3, as can be seen by numerical simulations of the SK model
\cite{BHAYO85}).

When we go in less than $6$ dimensions the quantity $g_{-}$ starts to 
be less than one and decreases with the dimensionality of the space.  
When, by decreasing $D$, the value of $g_{c}$, i.e.  the value of the 
Binder cumulant at the crossing point becomes close to $g_{-}$, the 
effect of crossing becomes very difficult to detect \cite{DOKO93}.  
One also expects that $g_{c}$ becomes a non trivial function of $D$ for 
dimensions lower than 6.

 This behavior is related to the lack of scaling in the mean field 
theory.  Indeed the function $P(q)$ can be written
as
\be
P(q)= \tilde{P} (q) + (1-x_{M}) \delta(q-\qea)
\ee
where the term $\tilde{P} (q)$ does not contain a delta function at $\qea$.  The 
quantity $x_{M}$ gives the probability of finding two different systems with an 
overlap $q < \qea $.  In mean field theory $x_{M}$ is proportional to $T_{c}-T$: 
since a pure number is proportional to the distance from the critical 
temperature scaling is badly violated.  On the other end it was shown
\cite{DOKO93} that in less than $6$ dimensions scaling is restored and the 
function $P(q)$ scales as

\begin{equation}
  q_{\rm EA} P(q)=f(\frac{q}{q_{\rm EA}})\ ,
  \protect\label{SCALING}
\end{equation}
where $q$ vanishes as a $|T-T_{c}|^{\beta}$.  
At least a partial verification \cite{DOKO93} of equation (\ref{SCALING}) has 
been done by verifying that near $T_{c}$ the quantity $q_{\rm EA} P(0)$ does not 
depend on $T$.

\subsection{Non-Zero Magnetic Field\protect\label{SS-D4FIEL}}

An important prediction of the mean field solution concerns the 
existence of a transition even for non zero magnetic field.  When the 
magnetic field is small enough it exists a $h$ dependent temperature 
$T_{\rm AT}(h)$ (the de Almeida-Thouless line) where the overlap 
susceptibility diverges.  Below the field dependent critical 
temperature the function $P(q)$ is non trivial.

It is difficult to study numerically the transition in field in good 
detail \cite{CAPAPASO90,CAPAPASO90B,CAPAPASO91,GRAHET}.  The function 
$P(q)$ is symmetric around the origin at $h=0$, and it is concentrated 
at positive $q$ values for non zero $h$.  If $h$ is too small and the 
volume is not too large, one finds a tail of configurations with 
negative $q$.  This tail disappears when increasing the volume, but 
complicates the analysis \cite{CAPAPASO90,CAPAPASO90B,CAPAPASO91}.  
This region is relevant for the cross-over behavior from $h=0$ to 
$h\ne 0$.  If $h$ is not so small (for example for an $h$ such to 
induce a magnetization of $0.15$), the critical temperature is 
decreased by a large factor as compared to the $h=0$ case (of circa 
$40\%$ in the specific case of $m=0.15$) and in this low temperature
region measurements are much more difficult .

The present data \cite{CIPARIRU93,PIRI} support the existence of a transition: 
at low temperatures the overlap susceptibility diverges roughly proportionally 
to the volume and the function $P(q)$ strongly fluctuates from system to system.  
Studies of the system in presence of an external field (conjugated to the 
overlap) which couples two replicas suggest the presence of discontinuities at 
$\epsilon=0$, but a relative large extrapolation is needed for reaching these 
conclusions.

Unfortunately for $h\ne 0$ the values of the various Binder cumulants (related 
to skewness and kurtosis) as a function of the temperature have a rather complex 
behavior, and it is not clear how to use them to locate the phase transition 
point.  Also the theoretical situation is very confused: the renormalization 
group predictions for the critical exponent cannot be computed because no fixed 
point has been found \cite{BRAYROB}.  The result is puzzling and no convincing 
interpretations have been yet presented.

We believe that a much more careful study of the properties at non
zero magnetic field above and below the De Almeida-Thouless line is
very important and the present situation can be strongly improved in
the next future.

\subsection{Out of Equilibrium Dynamics\protect\label{SS-D4DYNA}}

We will discuss here, again (see (\ref{SS-D3DYNA})), an out of 
equilibrium approach.  In some situations that can be very helpful (we 
will see that in the $4D$ case we can even measure $\qea$ by an 
off-equilibrium technique).  Here one measures the relevant quantities 
as a function of time.  Often they can be fitted extremely accurately, 
in a large time window, by power laws, i.e.  by a form $A+Bt^{-C}$: in 
this way, especially if the exponent $C$ is not too small, one can 
perform the $t\to\infty$ limit quite precisely.  One further advantage 
of the method is that one can work with very large lattices.  Taking a 
lattice size much larger that the dynamic correlation length allows 
to make finite size corrections very small.

In the following we will mainly focus on the relation between off
and on equilibrium regimes, by describing mainly the work of 
\cite{PARIRU96}. We will see that it will be possible to establish 
strong links between the two regimes.

Asymptotically an equilibrium situation is reached by making $t_w$ 
large in the correlation functions of (\ref{E-CORDYN}) (so that the 
system is at equilibrium on very large time scales), and then 
considering large, but small compared to $t_w$, measuring times $t$.  
In this case we can expect to find power like corrections to $\qea$.  
We can write

\be
  q(t) \equiv \lim_{t_w \rightarrow \infty}
  C(t,t_w) =  \left( q_{\rm EA} + a t^{-x} \right)\;\;\;\mbox{for}\;\;
  t\gg 1 \ .
  \protect\label{EA}
\ee
Here we are saying that if we wait a large time the system will be 
equilibrated on time scales smaller than the waiting time.  So if we 
measure correlations up to these scales we will find that the 
autocorrelation function tends to a plateau that is exactly the 
Edwards-Anderson order parameter: for $t\simeq t_w$ there will be a 
crossover, and the correlation function will decay to zero.  The most 
part of numerical simulations, done in a region of short waiting 
times, were dealing with this second regime 
\cite{OGMO85,OGIELSKI85,RIEGER93}, observing a power decay to $q=0$. 
Using a large waiting time has recently allowed \cite{PARIRU96} to 
clearly detect the effect implied by (\ref{EA}).

One uses a large $\frac{t_w}{t}\ge 32$ ratio. In these conditions the 
numerical data are well fitted by the form

\be
  C(t,t_w) =(q_{\rm EA} + a t^{-x})\ \overl{C}(\frac{t}{t_w})\ ,
  \protect\label{scaling}
\ee
where for $z\to 0$ one has $\overl{C}(z)\simeq1-c_1 z^\zeta$.  One 
first determines the rescaling function $\overl{C}(\frac{t}{t_w})$ by 
fitting the numerical data for the autocorrelation function at a fixed 
value of $t$ (as a function of $t_w$).  Then one divides away from the 
numerical data the value of $\overl{C}$: the fact that all the 
rescaled points, at different $t$ and $t_w$, fall on a single, 
universal curve, is a test of the fact that (\ref{scaling}) was a 
correct Ansatz.

\begin{figure}[htbp]
  \begin{center}
    \leavevmode
    \epsfysize=250pt\epsfxsize=300pt\epsffile{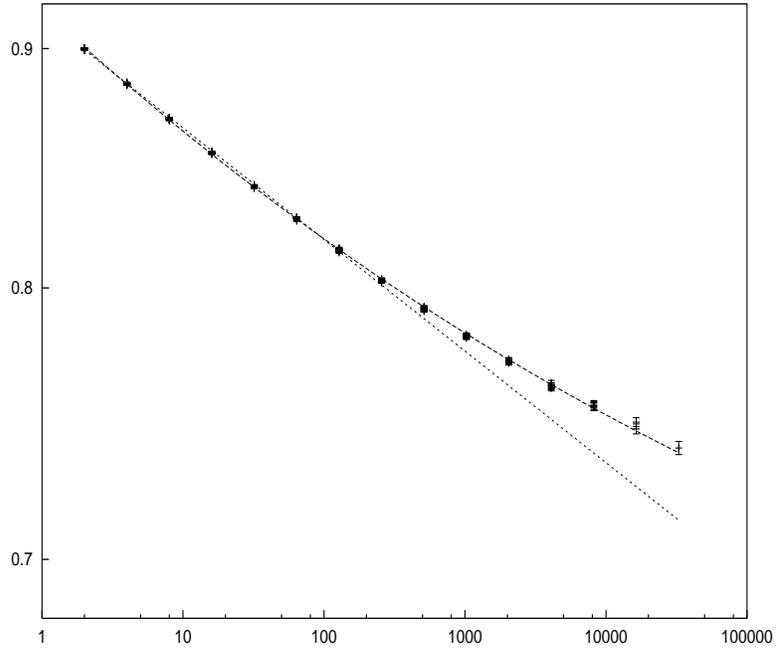}
  \end{center}
  \protect\caption[0]{$C(t,t_w)/\overl{C}(t,t_w)$ 
    versus $t$ at $T=0.9$. The dashed
    line is the pure best power fit while the solid line is the best power
    fit including a constant: 
    $(0.60\pm 0.04) +(0.32\pm 0.04)t^{-0.08\pm 0.01}$.
    From ref. \cite{PARIRU96}.}
  \protect\label{fig:q}
\end{figure}

After these steps one can try to fit the scaling curve to a power 
behavior.  The numerical date together with the fit are shown in 
figure (\ref{fig:q}).

It is clear from this figure that for large $t$ (but still in the 
regime $t_w/t\ge 32$), the data do not follow a pure power fit 
($t^{-x}$) and there is a correction that can be taken in account by 
fitting to the form $q_{\rm EA} +a t^{-x}$.  In figure \ref{fig:q} we 
also plot this second fit.

\begin{figure}[htbp]
  \begin{center}
    \leavevmode
    \epsfysize=250pt\epsfxsize=300pt\epsffile{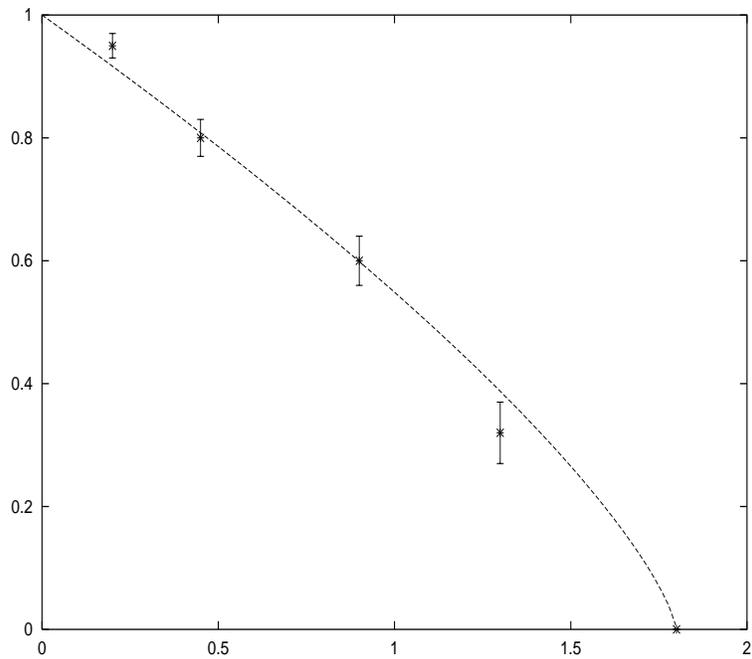}
  \end{center}
  \protect\caption{$4D$ Edwards-Anderson order parameter, computed 
    from non-equilibrium dynamics, versus $T$.
    From ref. \protect\cite{PARIRU96}.}
  \protect\label{fig:qea}
\end{figure}

The best estimates for $\qea$ as a function of $T$ are shown in  
(\ref{fig:qea}). The dashed line is the function

\be
  q_{\rm EA}(T)\simeq\left(\frac{T_c-T}{T_c} \right)^\beta .
  \protect\label{beta}
\ee
drawn using the values obtained by equilibrium simulations of the 
model \cite{PARIRU96}:
$T_c=1.8$ and $\beta=0.74$. The line is only a guide to the eye, but 
it coincides very well with the numerical data, even far from $T_c$ 
(where we do not expect a priori that a simple power decay holds).

\subsection{Ultrametricity\protect\label{SS-D4ULTR}}

Verifying the ultrametric structure of spin glass models by numerical 
simulations is a difficult task. Even for the SK model, where we know 
analytically what to expect, fully satisfactory numerical checks have 
not been yet obtained. Still, the question is very important: is the 
phase structure of finite $D$ models reminiscent of the ultrametric 
organization of the mean field solution? Cacciuto, Marinari and 
Parisi \cite{CAMAPA} have discussed this issue in the $4D$ case, and 
found a positive evidence, that we will discuss in the following.
The interested reader can read the interesting introductions and 
discussions of  \cite{RATOVI,PARULT}: mean field techniques allow
advanced computations about the ultrametric structure of the phase 
space \cite{FRPAVA,FRPAVC}.

A good introduction to ultrametricity for physicists is 
in \cite{RATOVI}. Here we just remind the reader that the usual 
triangular inequality

\be
  d_{1,3} \le d_{1,2} + d_{2,3}\ ,
\ee
is substituted in spaces endowed with an ultrametric distance by the 
stronger inequality

\be
  d_{1,3} \le \max \left( d_{1,2} , d_{2,3}\right)\ .
\ee
In an ultrametric space all triangles have at least two equal sides, 
that are larger or equal than the third side. An hierarchical tree is 
a very good way of representing an ultrametric set of states. In the 
solution of the mean field spin glass theory one finds an exact 
ultrametric structure: states are organized on an hierarchical tree, 
and if we pick up three equilibrium configurations of the system and 
compute their distance we find an ultrametric triangle.

Reference \cite{CAMAPA} is based on a constrained Monte Carlo 
procedure.  One updates three replicas of the system (in the same set 
of couplings), and constrains the distance of replica one and replica 
two to a given value $q_{1,2}$, and the distance of replica two and 
replica three to $q_{2,3}$ (that can be equal to $q_{1,2}$).  We have 
three replicas, two distances are fixed and we measure the third one, 
that we call $q$.  For example if one fixes both values to some 
fraction of $q_{\rm EA}$ (in the case of \cite{CAMAPA} to $\frac25 
q_{\rm EA}$) 
an ultrametric structure would imply that $q\ge \frac25 q_{\rm EA}$, 
while 
the usual triangular inequality would only imply that $q\ge -\frac75 
q_{\rm EA}$. Obviously the choice of the constraint is crucial to obtain 
a sharp difference from the usual situation of an Euclidean metric.

\begin{figure}
  \epsfysize=350pt\epsfxsize=250pt
  \centerline{\rotate[l]{\epsffile{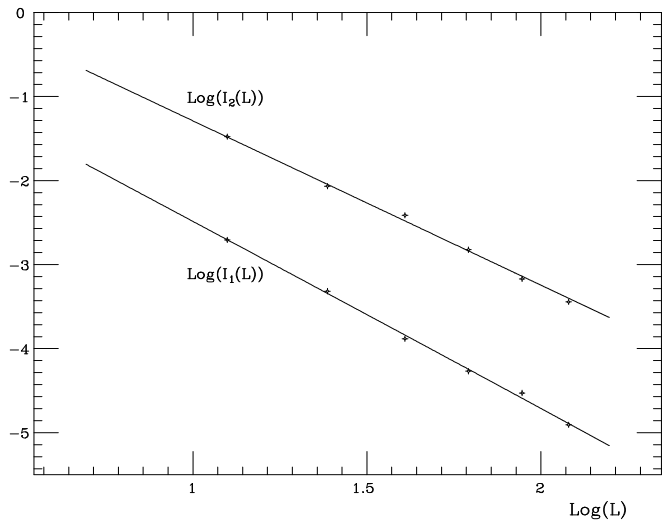}}}
  \protect\caption[1]{
    The integral $I^{L}$ as a function of $L$, in double log scale.
    The lower points
    are for the case where we have fixed 
    $q_{1,2}= q_{2,3}$, the upper points
    where $q_{1,2}\ne q_{2,3}$ (see the text).}
  \protect\label{F-ULTRAM}
\end{figure}

It has been possible to thermalize lattices of up to $8^4$.  The 
computation turns out to be, as we will see, very successful. The 
most 
serious problem turns out to be in the usual finite size effects: 
finite size effects are serious in spin glass models, and in this 
computation they appear clearly. In order to be more quantitative we 
define the integral

\be
  I^L \equiv 
  \int_{-1}^{q_{\rm min}} {\rm d}q~ \left( q(L)-q_{\rm min}\right)^2 \ P(q) \
  + 
  \int_{q_{\rm max}}^{+1} {\rm d}q~ \left( q(L)-q_{\rm max}\right)^2 \ P(q)\  ,
\ee
where $q_{\rm min}$ is the minimum $q$ allowed (for us, for example, 
$q_{\rm min}=q_{1,2}$), and $q_{\rm max}=q_{\rm EA}$. 
$I^L$ goes to zero if the 
system is ultrametric. We plot $I^L$ in fig. (\ref{F-ULTRAM}) for the 
two choices of the constraint that have been discussed in 
\cite{CAMAPA}.

For example in the case of two equal distances a very good best fit 
shown in the figure gives

\be
  I^L \simeq (-0.0001 \pm 0.0005) + (0.76 \pm 0.03) 
  L^{-2.21\pm0.04}\  .
\ee
It is remarkable that the mean field computations of 
\cite{FRPAVA,FRPAVC} give an exponent of $\frac83\simeq 2.67$, for 
the 
deviations from a pure ultrametric behavior in a finite system. Not 
only one finds a system that for large $L$ is converging to an 
ultrametric behavior, but the rate of the convergence is very similar 
to the one one can compute in the mean field model. This is one of 
the quantitative agreements that make the relation of the mean field 
solution and the finite dimensional models clear and impressive.

\section{D=2\protect\label{S-D2}}

We do not have enough space to enter in many details about the $2D$ 
case \cite{WANSWE,BERCEL,LIANG,KARDAR,RIEGER4,RIEGER5,RIEGER6,LEMCAM}.  
We will briefly discuss the statics of the problem, the out 
off-equilibrium dynamics and try to stress some important points, like 
the nature of the $T=0$ divergence.

\subsection{Statics\protect\label{SS-D2STAT}}

As we have discussed the original Bhatt and Young work \cite{BHAYO88} 
seems already to shed a clear light on the $2D$ cases (we will discuss 
in a few lines recent doubts \cite{KARDAR}).  For $J = \pm 1$ 
couplings one was finding a clear signature for a $T=0$ transition, 
with power law divergences with $\nu=2.6\pm 0.4$, $\eta=0.20\pm 0.05$
and $\gamma=4.6\pm 0.5$.

Recent transfer matrix calculations \cite{KARDAR}, mainly looking at 
the complex zero structure of the partition function, seem however to 
be opening doubts, supporting a correlation length that would be 
diverging exponentially (see also our discussions of section 
(\ref{S-D3})). So one would have that (for $J=\pm 1$ in $2D$) 
$\xi\simeq \exp(\frac{2}{T})$. The question does not seem to be 
solved at the moment.

\begin{figure}
  \epsfysize=270pt\epsfxsize=250pt
  \centerline{\epsffile{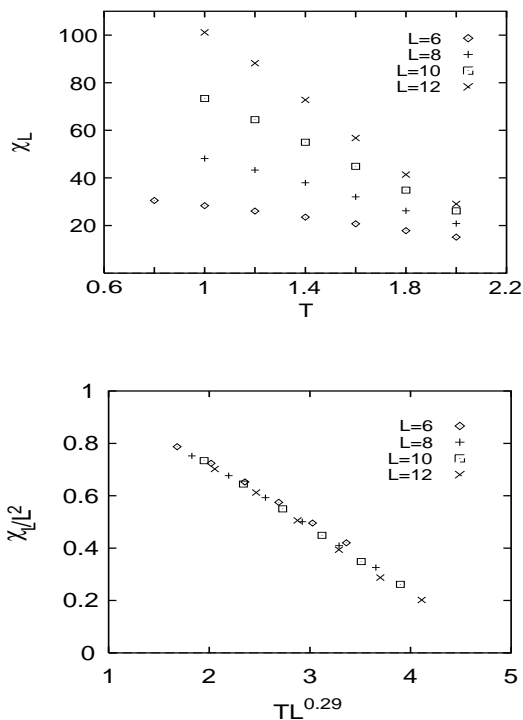}}
  \protect\caption[1]{$2D$ Gaussian $J$ spin glass:
   equilibrium values of the susceptibility 
   depending on temperature and system size. In the upper figure the 
   bare data, in the lower part the rescaled data.}
  \protect\label{fig:sus_res}
\end{figure}

The model with Gaussian couplings $J$ has been discussed in detail in 
\cite{RIEGER5}, under the two aspects of the $T=0$ structure and of 
the finite $T$ equilibrium. $T=0$ has been analyzed by determining ground 
states thanks to a branch and cut algorithm. Under this approach the 
authors find $y=-.281\pm 0.002$ (and since one expect 
$y=\frac{1}{\nu}$ that gives $\nu=3.56\pm.02$). A Monte Carlo 
simulation is used to determine the finite $T$ behavior. Assuming a 
$T=0$ power law divergence and using continuous couplings (with no 
accidental degeneracy) one has that $\eta=\beta=0$ and 
$\frac{\gamma}{\nu}=2$, leaving only one independent exponent, say 
$\nu$, to be determined. In fig. (\ref{fig:sus_res}) we show the 
susceptibility from ref. \cite{RIEGER5}, and the good finite size 
scaling behavior obtained by using $\nu=3.45$. 

By also using a detailed analysis of the Binder parameter $g$ one gets 
$\nu=3.6\pm 0.02$, in very good agreement with the $T=0$ result for 
$y$.  Ref.  \cite{RIEGER5} also give a quite precise estimate of the 
magnetization exponent $m_\infty(h)\simeq h^{\frac{1}{\delta}}$, 
$\delta = 1.48 \pm 0.01$ (there is a problem since one would expect 
$\delta=1-y$, that is not well verified by the data).  Also they study 
the chaotic behavior one expects in spin glasses.  Also a very recent 
paper \cite{RIEGER6} is based on $T=0$ exact ground states, and allows 
a determination of the stiffness exponent, that turns out to be small 
and negative, $-0.056\pm 0.006$.

At last we note that Lemke and Campell \cite{LEMCAM} have studied the 
$2D$ model with next-nearest neighbor interactions and found signs of 
the possible existence of a spin glass phase.

\subsection{Out of Equilibrium Dynamics\protect\label{SS-D2DYNA}}

We will give here a few details about the off equilibrium dynamics in 
the $2D$ model, by mainly following \cite{RIEGER4} and 
\cite{RIEGER96}.  The main points are maybe that interrupted aging can 
be observed in detail (since there is no phase transition the system 
eventually converges to a time translational invariant regime), and 
that again the predictions of the droplet model do not fit the 
numerical data.  As usual, the numerical studies are mainly based on 
the measurement of the correlation function defined in equation 
(\ref{E-CORDYN}).

\begin{figure}
  \begin{center}
    \leavevmode
    \epsfysize=400pt\rotate[r]{\hbox{\epsfbox[107 3 586 757]{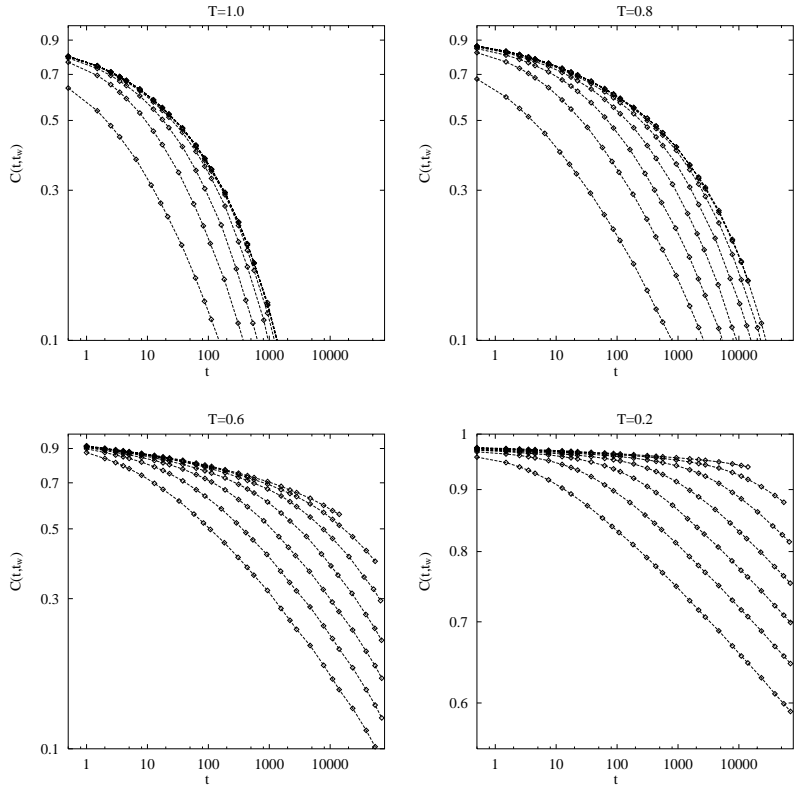}}}
  \end{center}
  \protect\caption{Autocorrelation function $C(t,t_w)$ as a function of
    time $t$ for $t_w=5^n$  ($n=1,\ldots,8$) at $T=1.0$ and $0.8$,
    ($n=2,\ldots,8$) at $0.6$ and $0.2$. The system size is $L=100$
    and the disorder average was performed over 256 samples. The error 
    bars are smaller than the symbols. From ref. \protect\cite{RIEGER4}.}
  \protect\label{fig:rieger_1}
\end{figure}

The first result, see figure (\ref{fig:rieger_1}) ), is that for 
waiting times $t_w$ larger than a given value $\tau_{\rm eq}$ the 
curves of the autocorrelation function, $C(t,t_w)$, as a function of 
$t$ for different $t_w$, collapse.  This implies that the system 
equilibrates.  One can identify $\tau_{\rm eq}$ as the time necessary 
to reach the equilibrium situation (the regime where the 
fluctuation-dissipation theorem holds).  This is what is called 
interrupted aging.  The equilibration time grows when the temperature 
decreases.  For lower temperatures the equilibration time 
becomes larger that the simulated time and the situation is not 
qualitatively different from the one three or four dimensions.

The correlation function, $C(t,t_w)$, follows empirically the
scaling law

\be
  C(t,t_w)=f(\frac{t}{\tau(t_w)})\ ,
\ee
where $f(x)$ is a scaling function and the time scale, $\tau(t_w)$, is 
proportional to $t_w$ as $t_w \ll \tau_{\rm eq}$, and reaches a 
plateau when $t_w > \tau_{\rm eq}$.  In the latter regime the variable 
of the scaling function will be $\frac{t}{t_w}$.  The droplet model 
suggests a dependence over $\frac{\log(t)}{\log(t_w)}$ that is clearly 
unable to describe the data.

Also measurements of the correlation length $\xi(t_w)$ give precise 
results.  The fit to a pure algebraic behavior, $\xi \propto 
t^{\alpha(T)}$, with $\alpha\simeq 0.2 T$ works well.  The droplet 
approach predicts $\xi(t_w) \propto (\log t_w) ^{1/\psi}$, that here 
also gives a reasonable fit, with $\psi = 0.65 \pm 0.01$ independent 
of $T$.

\section*{Appendix 1: On the Definition of Pure States.\protect\label{S-PURE}}

We will give here a few more details about the problem of defining 
{\em pure states}.  We will use this notion in a physical way, which 
may be different from the approach used by the mathematical physics 
community.

The basic idea is rather simple.  Let us consider for simplicity a 
spin system with nearest neighbor interaction on the lattice.  
Everything works fine for an {\em actually infinite} system.  We 
define a state $\rho(C)$ as a probability distribution over the 
configurations $C$ of the {\em infinite} system \footnote{We use here 
and in the following an informal language: all what we are saying can 
be phrased in a precise mathematical language, but such a 
reformulation would be out of place here.}.  A state is said to be a 
local equilibrium state (or a DLR state \cite {RUELLE}) if the 
restriction to a finite volume of the probability distribution that 
characterizes it is given by the Boltzmann formula.

A theorem says \cite {RUELLE} that any DLR state can be decomposed as 
the sum, with non negative coefficients, of pure DLR states:

\be
  \langle \cdot \rangle =\sum_{\al} W_{\al} \langle \cdot \rangle_\al\ .
  \label{DECO}
\ee
Pure states are the ones for which the only possible decomposition has 
one $W_\gamma=1$ and all the other weights equal to zero.  In other 
words the DRL states are a convex set and the pure states are the 
extremal states of this set.  The pure states can also be 
characterized by the clustering property: in pure states the connected 
correlations functions go to zero at large distances, or equivalently 
in pure states intensive quantities do not fluctuate \cite{KASTLER,KASROB} .

The proofs which are needed are very simple \footnote{The only tricky 
point is to prove the clustering property for pure states.} if one 
uses the appropriate mathematical setting \cite{KASTLER}.  Hard 
problems start when we have to show that this nice construction is not 
empty, i.e.  when we have to prove that local equilibrium states do 
exist for the infinite system.  The simplest way we have to accomplish 
this task is to take a finite volume system and to show that the 
infinite volume limit of the Boltzmann Gibbs probability does exist 
and it is a local equilibrium state.  In this construction there is 
the freedom to chose the boundary conditions of the system, that could 
lead to different local equilibrium states.  If the boundary conditions 
are chosen in an appropriate way (e.g. all spins up in a ferromagnet) 
a pure state is obtained.

This decomposition into pure states is well known.  It was developed 
thirty years ago for the case of translational invariant Hamiltonians 
\cite{KASTLER}. In the case of spin glasses (and more generally of 
other system with quenched non translationally invariant disorder) 
things are much more difficult.  The very concept of a probability 
distribution over configurations of the {\em actually} infinite 
system needs extreme mathematical care.  Just consider the example of 
a ferromagnet at low temperature in presence of a random quenched 
magnetic field.  We know that for a {\em finite}, large system, there 
is a magnetization which is equal to $\pm 1$, the sign being the one 
of $h_T\equiv\sum_i h_i$, provided that $h_T^2$ is a quantity of order 
of the volume (as usually happens).  Everything is clear!  However if 
we want to consider an actually infinite system which is the sign of 
$h_T$?  We could consider the function $s(L)\equiv \sign\sum_{i=-L,L} 
h_i$, but this leads nowhere because if the $h_i$ are random variables 
with zero average, $s(L)$ does not have a limit when $L$ goes to 
infinity.

The real problem with spin glasses and with other disordered systems 
is that it is extremely difficult to control the Boltzmann Gibbs 
probability in the infinite volume limit.  The previous example of a 
ferromagnet in a random field strongly suggests that such limit may 
not exist, at least not in a naive way.  Similar conclusions are valid 
for spin glasses in the mean field approach \cite{MEPAVI}, and they 
have been conjectured to be valid also for short range glasses.  
Sometimes one refer to this phenomenon as chaotic dependence of the 
properties of the system on the size \cite{NS1}.  To deal with this 
problem different techniques have been suggested (for a recent 
discussion see reference \cite{NS3}).  Using different definitions 
leads to different results, that potentially describe very different 
physical pictures \cite{NS1,NSREP}.

A decomposition into pure states of the Boltzmann Gibbs probability 
distribution for an infinite system is only possible if the Boltzmann 
Gibbs probability distribution exists in the infinite volume limit and 
this does not seem to be the case of many disordered systems.  An 
alternative approach consists in making an approximate decomposition 
into pure states for a finite system; this decomposition must coincide 
with the usual definitions in the case where the infinite volume limit 
can be done without difficulties (i.e.  where there is no chaotic 
dependence on the side).

Let us see how one could define approximate pure states in a large but 
{\em finite} system.  In this way we are giving a different, but maybe 
more physical, definition of a state.

Let us consider a system in a box of size $L$.  We partition the 
configuration states in regions, labeled by $\al$, and we define the 
averages restricted to these regions \cite{PAR1,PAR2}.  We have to 
impose that the restricted averages on these two regions are such that 
connected correlation functions are small at large distance $x$, i.e.  
they go to zero faster than a given function $A(L)$ such that $\lim_{L 
\to \infty}A(L)=0$.  In this way we recover eq.  (\ref{DECO}) for a 
finite system.  In the case of a ferromagnet the two regions are 
defined by considering the sign of the total magnetization.  There are 
ambiguities with those configurations which have exactly zero total 
magnetization, but the probability that such a configuration occur is 
exponentially small at low temperature.

Physical intuition tells us that this decomposition can be done (at 
least for familiar systems), otherwise it would make no sense to speak 
about the spontaneous magnetization of a ferromagnetic sample or to 
declare that a finite amount of water (at the melting point) is in the 
solid or liquid state (also all numerical simulations gather data that 
are based on these kinds of notions, since systems we can store in a 
computer are always finite).  We strongly believe that these 
statements do make sense, although their translation in a rigorous 
mathematical setting has never been done (as far as we known) also 
because it is much simpler (and in many cases sufficiently enough) to 
work directly in the cozy infinite volume setting.

We assume that such decomposition can be done also in spin glasses 
(the contrary would be highly surprising for any system with a short 
range Hamiltonian).  Therefore the {\em finite} volume Boltzmann Gibbs 
measure can be decomposed in sum of the finite volume pure states 
according to the previous definitions.  The states of the system are 
labeled by $\al$ and they satisfy eq.  (\ref{DECO}).  The function 
$P(q)$ for a particular sample is given by

\be
\sum_{\al,\beta} W_\al W_\beta \delta(q_{\al,\beta}-q)\ ,
\ee
where $q_{\al,\beta}$ is the overlap among two generic configurations 
in the state $\al$.

This definition of states is used only at a metaphorical level.  The 
predictions of the mean field theory concerns correlation functions 
computed in the appropriate ensemble \cite{BREAKS} and computer 
simulations measure directly these correlation functions.  The 
decomposition into states (which is never done explicitly during 
computer simulations) is an interpretative tool which describes the 
complex phenomenology displayed by the correlation functions in a simple 
and intuitive way.  We could alternatively define the function $P(q)$ 
as

\be
\int dq P(q)q^s ={\sum_{i,k=1,N} <\si_i \si_k>^{2s} \over N^2}\ ,
\ee
but this definition would have much less intuitive appeal the previous one.

The two approaches, the replica analysis of the finite volume 
correlations functions (and the results which can be stated in a 
simple and intuitive way by using the idea of decomposition into 
states of the Boltzmann Gibbs measure) and the construction of pure 
states for the actually infinite system, give complementary 
information which can be hardly compared one with the other.  In the 
replica method one obtains information only on those states whose 
weight $w$ does not vanish in the infinite volume limit \footnote{As 
it stands this sentence may be misleading because it could seem to 
describe the property of a given same state when we change the volume.  
A more precise (and also heavier) formulation is the following: for 
each particular volume the replica method gives information on the 
states (defined for that particular model) whose weight $w$ is not too 
small when $N$ is very large.}.  All local equilibrium states have the 
same free energy density; however the differences in the total free 
energy may grow as $L^{(D-1}$.  From an infinite volume point of view 
all these states are equivalent, from a finite volume point of view 
only the state with lower free energy and the states whose total free 
energy differ from the ground states by a finite amount are relevant.

For example in the ferromagnetic case (in more than two dimensions at 
sufficient low temperature) there are equilibrium states which have in 
half of the infinite volume positive magnetization and in the other 
half negative magnetization.  These states are invisible in the 
replica method because their weight (when restricted to a finite 
volume system) goes to zero as $\exp(-A \L^{D-1})$ (special 
techniques, i.e.  coupling replicas may be used to recover, at least 
partially, this information).  In the replica method the states are 
weighted with the corresponding Boltzmann Gibbs weight and this weight 
can be hardly reconstructed from an analysis done directly at infinite 
volume.

\section*{Appendix 2: Simulated Tempering\protect\label{S-TEMPER}}

In this section we will describe the so called tempering methods 
\cite{MAPA93} (see also the lecture notes in \cite{MARINARI96}).  In 
these methods the temperature becomes a dynamic variable.  In 
particular we will describe the simulated tempering method 
\cite{MAPA93} and a crucial variation, the powerful parallel tempering 
scheme \cite{HUNE95,TESI}.  The multicanonical methods 
\cite{BERCEL,BECEHAN94} have very similar roots, and can be also 
employed very effectively, but we will not describe them here.
These methods has been used to simulate very effectively 
a wide range of physical problem (see \cite{MARINARI96} for a list).

The basic idea of both methods is to move in the temperature space 
(always staying at thermodynamical equilibrium with respect to a 
suitable probability distribution) to avoid being trapped fro high 
energy barriers: the system change its temperature, goes up to the 
paramagnetic phase and eventually goes back to the lower temperatures.  
With high probability in different visits the system will visit new 
local minima (if the phase space has a reasonable shape).

Let us introduce the tempering scheme.  We have the original phase 
space, that we will denote by \{X\}, a Hamiltonian $\cH (X)$ and a 
{\it new} variable $m$ which takes $M$ values ($\{m\}=\{1,...,M\}$).  
We extend the original phase space to a new space $\{X\} \times 
\{m\}$.  The probability for a element, $(X,m)$, of this extended 
phase space to occur is given by

\be
  P(X,m)\equiv \frac{1}{\cZ_{\rm EXT}}
  \exp\left[-\cH_{\rm EXT}(X,m)\right]\ ,
\ee
where

\be
  \cH_{\rm EXT}(X,m)\equiv \beta_m \cH(X)-g_m\ ,
\ee
and

\be
  \cZ_{\rm EXT}\equiv\sum_{m=1}^M \sum_{\{X\}}
  \exp\left[-\cH_{\rm EXT}(X,m)\right]=
  \sum_{m=1}^M e^{g_m} \cZ(\beta_m)\ .
\ee
The extended partition function is the weighted sum of the $M$
partition functions ($\cZ(\beta_m)$) at given $\beta_m$, and

\be
  \cZ(\beta_m)\equiv \sum_{\{X\}} \exp\left[-\beta_m \cH(X)\right]\ .
\ee
The $\beta_m$ are dynamic variables which will be allowed to span a 
set of given values (e.g.  the inverse temperatures that we want to 
simulate) and the $g_m$ must be fixed before the run begins.

If we fix $m$, it is obvious that the probability distribution for $X$
is given by the usual Boltzmann weight with $\beta=\beta_m$. Moreover,
the probability to find a given value of $m$ is

\be
  P(m) \equiv \sum_{\{X\}} P(X,m)= \frac{\cZ(\beta_m)
  e^{g_m}}{\cZ_{\rm EXT}} = \frac{1}{\cZ_{\rm EXT}}
  \exp(-\beta_m f(\beta_m) +g_m)\ ,
\ee
where  $f(\beta_m)$ is the free energy at fixed $m$ (i.e.
$\beta_m f(\beta_m)=-\log \cZ(\beta_m)$).

If we choose $g_m=\beta_m f(\beta_m)$ all the different
$m$'s have the same probability, equal to $1/\cZ_{\rm EXT}$.
In this case $\cZ_{\rm EXT}=M$. 

Now, we will compute the probability of jumping between two 
consecutive inverse temperatures $\beta_m$ and $\beta_{m+1}$ (we are 
assuming that the $\beta$'s are ordered: $\beta_m < \beta_{m+1} < 
\beta_{m+2} < \ldots $).  The variation of the extended Hamiltonian 
for a given configuration $X$ is

\be
  \Delta \cH_{\rm EXT}= E_{\rm inst} \delta -(g_{m+1}-g_m)\ ,
\ee
where $\delta\equiv \beta_{m+1}-\beta_m$ and $E_{\rm inst}$ is the 
instantaneous energy, $E_{\rm inst}\equiv \cH(X)$.  Expanding 
$g_{m+1}=\beta_{m+1} f(\beta_{m+1})$ near $\beta_m$ we obtain

\begin{eqnarray}
\nonumber
g_{m+1}\equiv g(\beta_{m+1})&=&g(\beta_m) +
\left .\frac{d g(\beta) }{d\beta} \right|_{\beta=\beta_m}\delta\\
&+&\frac{1}{2}
\left .\frac{d^2 g(\beta) }{d \beta^2} \right|_{\beta=\beta_m}
\delta^2 +O( \delta^3)\nonumber \\
&=&E(\beta_m) \delta +\frac{1}{2} C(\beta_m) \delta^2 +O( \delta^3)\ ,
\end{eqnarray}
where $E(\beta_m)$ is the mean energy at $\beta_m$, $d g(\beta)/d
\beta=E(\beta)$ and $d E/d\beta=C(\beta)\equiv \langle \cH^2 \rangle
-\langle \cH \rangle ^2$.

By assuming that $E_{\rm inst}$ is close to $E(\beta_m)$, the 
variation $\Delta \cH_{\rm EXT}$ will be not large if we keep 
$C(\beta_m) \delta^2=O(1)$.  In this case we will have a reasonable 
acceptance ratio for the $\beta$ swaps.  This condition of $\delta$ is 
equivalent to impose that the energy histograms at $\beta_m$ and 
$\beta_{m+1}$ overlap.

At the critical point the specific heat ($C(\beta)$) diverges as

\be
  C(L,\beta_c) \propto L^{\alpha/\nu+d}\ ,
\ee
such that the condition on $\delta$ reads

\be
  \delta \propto L^{-(d+\alpha/\nu)/2}\ ,
  \label{ldelta1}
\ee
while in the non critical region $C(L,\beta)$
diverges with the volume, $L^d$, and

\be
  \delta \propto L^{-\frac{d}{2}}\ .
  \label{ldelta2}
\ee
The procedure used in the tempering method is composed by two steps (we
start the update from $ (X, \beta_k ) $):

\begin{enumerate}

  \item{We update the spin configuration $X$ to $X^\prime$ using, for 
  instance, the Metropolis or Heat Bath method at fixed $\beta_k$.  We 
  can repeat this step a certain number of times before going to 
  the next phase.}

\item{We try to update the inverse temperature $\beta_k$ to
  $\beta_{k\pm1}$ using a Metropolis like test: if $\Delta \cH_{\rm
  EXT}< 0$ we accept the change, otherwise we accept the change with
  probability $\exp(-\Delta \cH_{\rm EXT})$.}

\end{enumerate}

This procedure satisfies detailed balance.  From the previous 
discussion it should be clear that the most difficult part of the 
method is to fit the $g_m$ to the values of the free energies (on the 
contrary selecting the $\beta$ set is not a very demanding task).  This 
can be done by using an iterative procedure inside the simulating 
program: we change at run time the $g_m$ values until we obtain an 
uniform probability for the different $\beta$'s.

A typical run done using this method consists in:

\begin{enumerate}

\item Run a simple Metropolis algorithm in order to get
a first calculation of the free energies.

\item Run the simulated tempering and change, at run
time, the previous values of the free energies in order to obtain a
constant probability on $\beta$'s. 

\item Run the equilibrium simulations, with fixed $g_m$, and measure 
the interesting observables.

\end{enumerate}

\section*{Appendix 3: Parallel Tempering\protect\label{S-PARTEM}}

A great improvement to the previous method is the parallel tempering 
method (PT) \cite{HUNE95,TESI}.  The great advantage is that in this 
case we do not need to compute the partial free energies.  In the 
tempering method we have only had one system and a set of $M$ 
temperatures: the spin system was changing its $T$ value.  In the PT 
method we have $N$ system and $N$ $\beta$'s: we will try to swap the 
configurations with two different temperatures.  So, we will always 
have a system in a given temperature of our set. 

Now we have $N$ inverse temperatures $(\beta_1, \ldots, \beta_N)$ and $N$ 
non-interacting real replicas: the phase space is given by 
$\{X\}=\{X_1\}\times \ldots \times \{X_N\}$.  The partition 
function of the system reads

\be
  {\cZ}_{\rm EXT}=\prod_{i=1}^N \cZ(\beta_i)\ ,
\ee
and, as usual,

\be
  \cZ(\beta_i)=\sum_{\{X_i\}} \exp\left[-\beta_i \cH (X_i)\right]\ .
\ee
In the PT method the new phase space is the direct product of the
replicated original ones while in the tempering one it is the direct sum
(that is why we needed weights for the different terms of the sum).

For a given set of $\beta$'s, $(\beta_1,...,\beta_N)$, the probability 
of picking a configuration $X=(C_1,...,C_N)$ is

\be
  P(X;\beta_1,...,\beta_N)=\frac{1}{{\cZ}_{\rm EXT}}
  \exp\left[-\sum_{i=1}^N\beta_i \cH (C_i)\right]\ .
  \label{pro_pt}
\ee
We will define a Markov process for this extended system.  To do this 
we need to define a transition probability matrix $W(X,\beta; 
X^\prime, \beta^\prime)$ (that is the conditioned probability to 
exchange $X$ and $X^\prime$ without changing the $\beta$'s: i.e.  
initially we have two system $(X, \beta)$ and $(X^\prime, 
\beta^\prime)$ and we try to change to the situation $(X^\prime, 
\beta)$ and $(X,\beta^\prime)$).  The detailed balance condition for 
this system reads

\begin{eqnarray}
\nonumber
P
(\cdot \cdot \cdot,&X&,\cdot \cdot \cdot, X^\prime, \cdot \cdot \cdot;
\cdot \cdot \cdot, \beta,\cdot \cdot \cdot, \beta^\prime, \cdot \cdot 
\cdot)
W(X,\beta; X^\prime, \beta^\prime)\\
&=&
P(\cdot \cdot \cdot, X^\prime,\cdot \cdot \cdot, X, \cdot \cdot
\cdot;
\cdot \cdot \cdot, \beta,\cdot \cdot \cdot , \beta^\prime, \cdot\cdot 
\cdot)
W(X^\prime,\beta; X, \beta^\prime) \ .
\end{eqnarray}
Using equation (\ref{pro_pt}) we finally obtain

\be
\frac{W(X,\beta; X^\prime, \beta^\prime)}{W(X^\prime,\beta; X, 
\beta^\prime)}
=\exp(-\Delta)\ ,
\ee
where

\be
\Delta=(\beta^\prime-\beta)(\cH(X)-\cH(X^\prime)).
\ee
We can use a Metropolis like test: if $\Delta <0$ we
accept the change, otherwise we update with probability
$\exp(-\Delta)$.

The procedure for the PT  method is then:

\begin{enumerate}

\item Update independently the $N$ replicas using a standard MC
method simulating the usual canonical ensemble.

\item{Try to exchange $(X,\beta$) and $(X^\prime,\beta^\prime)$.  Accept 
the change if $\Delta <0$ and, if $\Delta >0$, change with probability 
$\exp(-\Delta)$.  Reject otherwise.}

\end{enumerate}

It is possible to show that $\delta\equiv \beta_{m+1}-\beta_m$ scales 
exactly like in the tempering method (see (\ref{ldelta1}) and 
(\ref{ldelta2})).


\end{document}